\documentclass[12pt]{JHEP3}

\usepackage{amsmath}

\title{Entropy current and equilibrium partition function in fluid dynamics}
\author{Sayantani Bhattacharyya$^{a,b}$\\
$^a$Ramakrishna Mission Vivekananda University,
Belur-Math, Howrah-711202\\
$^b$Indian Institute of Technology Kanpur, Kanpur-208016\\
email: sayanta@iitk.ac.in
 
}

\abstract{Existence of an entropy current with non-negative divergence puts a lot of constraints on the transport coefficients of a fluid, so does the existence of equilibrium. In all the cases we have studied so far we have seen an overlap between these two sets of constraints. In this note we shall try to explore the reason for such an overlap. We shall also see how a part of the entropy current could be determined from the equilibrium partition function.
}
\preprint{}

\begin{document}
\section{Introduction}\label{sec:introprelim}
Fluid dynamics is an effective description for  systems that are locally in thermodynamic equilibrium. For a fluid system, in a sufficiently small patch of space-time, we can always associate some local temperature, energy density and other thermodynamic variables.  These properties may vary in space and time but in such a slow fashion that the concept of local thermodynamic equilibrium is valid always.
These local thermodynamic properties are the basic variables of fluid dynamics.  We can choose them to be temperature $T(x)$, chemical potentials $\mu_i(x)$ and four-velocity $u^\mu(x)$.
The equations of fluid dynamics are simply the conservation equations for the stress tensor and the conserved currents of the system.
 In fluid dynamics the basic input is the constitutive relations which are expressions for the stress tensor and the conserved current in terms of the fluid variables.

Since the fluid variables are slowly varying with respect to space-time, we can treat the dynamics of fluid in a derivative expansion. Constitutive relations are also expanded in terms of the derivative of the fluid  variables. By naive symmetry analysis we could count the number of independent terms possible in each order in derivative expansion. Every independent term should be multiplied by some unknown coefficient, called the transport coefficients \cite{landau}.

 It is very difficult to compute the transport coefficients from the microscopic theory and therefore it is useful to constrain their number and structure from several other physical requirement that a consistent theory must satisfy.\\
\\
One such requirement is the local version of the second law of thermodynamics. According to the second law of thermodynamics total entropy should always increase for any evolution from one equilibrium to another equilibrium configuration. For a local theory like fluid dynamics we expect the entropy to increase locally. In other words there should exist an entropy current whose divergence is always positive definite for every non equilibrium fluid flow, consistent with fluid equations. We shall call this condition as `entropy condition'.

It turns out that just the existence of such an entropy current restricts the number and the structure of the transport coefficients significantly \cite{RomatImp}, \cite{son}, \cite{oddloga}, \cite{secondorder}, \cite{oddamos}, \cite{superfluident}, \cite{rindlerhydro},\cite{LifshitzOz2}. We can classify the constraints arising from `entropy condition' into two different categories. The expected ones are the `inequalities' for the values of different transport coefficients.  The others are the `equality-type' constraints. These are the ones  that genuinely reduce the number of transport coefficients by equating several of them which otherwise look independent from naive symmetry analysis.\\
\\
Like the local version of second law, the existence of equilibrium is another physical requirement\cite{equipart}, \cite{equipartamos}. More precisely we expect that fluid dynamics, when studied on a time independent background must admit atleast one time independent equilibrium solution. Also we should be able to generate the stress tensors and the charge currents of this equilibrium fluid from some partition function constructed out of the background metric and the gauge field. We shall refer to this condition as `equilibrium condition'. This condition also is not guaranteed by naive symmetry of the system and imposes several constraints on the transport coefficients \cite{rindlerhydro}, \cite{equipart}, \cite{equipartamos}, \cite{equitarun},   \cite{equiloga}, \cite{equikristan}, \cite{equisuperfluid}, \cite{equijustin}, \cite{complaint1}, \cite{complaint2}. 
\\
So far in every cases that have been analysed, we have seen that the `equality-type' constraints arising from `entropy condition' are same as the constraints imposed by `equilibrium condition'   \cite{oddloga}, \cite{oddamos}, \cite{superfluident}, \cite{rindlerhydro}, \cite{equipart}, \cite{equipartamos},  \cite{equiloga}, \cite{equisuperfluid}. In this note we shall explore the possible reason for such an equivalence in a general context. Eventually we shall show the following.
\newline
{\it If the `equilibrium condition' is satisfied by some constitutive relation, then we should always be able to construct an entropy current whose divergence is non negative for any  consistent fluid flow provided some transport coefficients (ones that are unconstrained by the `equilibrium condition') satisfy some further inequalities.}\\

In order to establish this, we shall develop an  explicit algorithm for the construction of an entropy current, starting from the equilibrium partition function of the system. Below we shall briefly describe how this construction works.

It is obvious that the equilibrium partition function will constrain the form of the entropy current to some extent. This is because 
once the partion function is known, using the rules of thermodynamics we can always compute the total entropy of the system in equilibrium. On the other hand, by definition, the integration of the time component of the entropy current over any space-like slice produces the total entropy of the system. Hence the time component of our entropy current, in equilibrium,  must reduce to what we get from the partition function.   As expected, this condition only  partly fixes the entropy current since those terms that vanish in equilibrium could never be fixed this way \cite{equipart},\cite{equisuperfluid}.

However, in the most general form of the entropy current, it is certainly possible to have many more terms which vanish when evaluated on equilibrium. In order to fix them we now turn to the study of the system in a time-dependent background. But we shall assume that the background changes with time adiabatically, i.e. the time variation is so slow that we shall keep track of only upto first order in time derivative. Now an entropy current, whose divergence is always non-negative, must be conserved for such adiabatic processes since the divergence could have any sign if we are considering only the first order time derivatives \cite{landau}. Many terms that could have been in the entropy current, but vanish in equilibrium, are non zero in this adiabatic situation.  They naturally get fixed in terms of the partition function once we demand that the entropy current is conserved also in case of this adiabatic variation of the background upto the appropriate order in space-derivatives. In addition to this, it turns out that we could construct an adiabatically conserved entropy current  if and only if  the equilibrium stress tensor and currents of the system are generated from the same  partition function, we are using to construct the entropy current. Therefore at this step we show that the existence of an equilibrium along with the partition function is a necessary condition for  local production of entropy in every fluid flow consistent with equation of motion. In other words it explains why the constraints (on the transport coefficients) arising from the `equilibrium condition' are always a subset of the constraints arising from the  `entropy condition'. \\

Now from the explicit examples we already know that the existence of equilibrium, though necessary, is not sufficient for the existence of an entropy current with non-negative divergence. There are also some `inequality type' constraints that we have to impose on the transport coefficients in order to ensure the `entropy condition'. The positivity of the viscosity or the diffusion constant are the examples. But working out specific examples at some given orders generally do not tell us how universal these inequalities are or whether we need  new  inequalities at every order in derivative expansion to satisfy the `entropy condition'. To understand this aspect, we next turn to the study of the system away from  adiabatic variation. We first compute the divergence of the entropy current, we  have already determined using equilibrium  partition function and its adiabatic variation. This time the divergence is computed exactly without ignoring terms that are zero under adiabatic approximation. In fact by construction each term in this divergence will necessarily have more than one time derivatives. Its form will depend on the coefficients of the partition function as well as the dissipative transport coefficients (i.e. the transport coefficients whose effects vanish in strict equilibrium). The transport coefficients enter the expression of divergence through equation of motion.

Unlike the adiabatic stuation, now we do not have any conservation equation to impose, rather only an inequality that the divergence is non-negative.  So it is natural that at this stage we could only generate some inequlaities for the dissipative transport coefficients (ones that appear in the expression of the divergence). The naive guess would be that at every order in derivative expansion there is some new inequality that we have to impose. But we shall show that only the first order dissipative transport coefficients are the ones that have to be constrained by inequalities to ensure the `entropy condition', all the rest are completely unconstrained. This we could show order by order in derivative expansion by modifying the entropy current further with adding new terms that are zero even in adiabatic approximation. The sole effect of these new terms are to absorb the contribution of the higher order dissipative terms into a positive definite form in the expression of the divergence. We give an explicit algorithm for how to do this showing that it could always be done.
\newline
So finally we have shown that if equilibrium exists and the first order transport coefficients have appropriate signs, the local production of entropy in every consistent fluid flow is gauranteed.

The result we found on inequalities could also be explained physically in terms of the stability of the equilibrium.
We know that when we do a linearized analysis of small fluctuations, the dissipative transport coefficients control the time evolution of the system. The appropriate sign of the transport coefficient is important in this matter as it ensures that the fluctuations eventually dissipate out and thus maintaining the the dynamical stability of the equilibrium. Also in derivative expansion, it is only the leading dissipative terms (in our case the first order coefficients like viscosity etc.) that will dominate the dissipation and as long as they satisfy the appropriate inequalities, the equilibrium is dynamically stable irrespective of the nature of the higher order coefficients.
Therefore the result we found in this note could also be re-worded as follows.
\newline
{\it Whenever there exits a dynamically stable equilibrium, entropy will always be locally produced for every fluid flow consistent with equations of motion.}\\

The organization of this note is as follows. First in section (\ref{sec:intro}) we shall roughly sketch our argument with a little more mathematical details.
Next in section (\ref{outline}) we shall explain our set up and the notation. In section (\ref{sec:general}) we shall explain how to construct $\hat J^\mu$, a current formed purely out of the background and treated as a precursor to the full fluid entropy current. In section (\ref{sec:fluident}) and section (\ref{sec:explicit})
we schematically describe how we can extend $\hat J^\mu$ to the full entropy current $J^\mu$.  In section (\ref{sec:ambiguity}) we shall discuss about the ambiguities involved in this prescription. In section (\ref{sec:example})  we shall implement this prescription for a very simple partition function to explicitly see how it works. Finally in section (\ref{sec:conclude}) we conclude.

In appendix(\ref{app:intarg}) we shall give some more arguments in support of the statements made in section(\ref{sec:explicit}).
In appendix(\ref{derivation}) we derive some of the equations used in the main text. In appendix(\ref{app:solution}) we shall solve some simple equations in a background with very slow time dependence to motivate one main assumption we used throughout this note.

All the analysis we have done here are for a four dimensional, parity preserving fluid with a single conserved abelian charge. But it can be easily generalized to multiple charges and other dimensions at least for parity-even cases. If the current and the stress tensor are anomalous, then there might be subtlety involved in determining the entropy current. See \cite{loga1}, \cite{loga2} for some recent analysis in this direction.

\section{Rough sketch of the method}\label{sec:intro}
In this section, we shall give a rough sketch of our construction with a little more mathematical  detail. The method will have two parts. In the first part, using the equilibrium partition function of the system we shall determine an entropy current which is conserved in equilibrium and adiabatically. In the next step we shall extend this current in a way so that at every order in derivative expansion its divergence is non-negative.

So we shall start with the most general background with a time-like Killing vector, which we identify with our `time-direction'. In such a background the system will be described by its partition function. From the partition function of the system,  we  calculate its total entropy, which is also time-independent to begin with. 
\newline
Now suppose, the background starts changing with time very slowly (so that we can do an independent expansion in time derivatives where $\partial_0^2$ is always negligible compared to $\partial_0$). The total entropy inside the bulk will also change. However in the leading order  in `$\partial_0$ expansion', algebraically the change could have either sign.
\newline
 But this will violate the second law of thermodynamics. Therefore at leading order the change in the net entropy must vanish. In other words the entropy change inside the bulk must be compensated by the entropy entering the region through some boundary current.

In terms of equation this means the following.
\newline
 If $\int_{space} \hat J^0$ is the total entropy (as derived from the equilibrium partition function) then
\begin{equation}\label{physeq}
\partial_0 \int_{space} \hat J^0 = -\int_{boundary} \hat J^i + {\cal O}(\omega^2)
\end{equation}
 where $\omega$ is the  frequency\footnote{Here each dimensionful quantity is measured in  the units of the equilibrium temperature $T_0$} of time dependence for the background satisfying $\omega^2\ll\omega$. 
 LHS of equation \eqref{physeq} could be determined using the partition function and usual rules of thermodynamics. We shall explicitly see that whenever the equations of motion and the `equilibrium condition' is satisfied, the LHS will be equal to a pure boundary term. From this  boundary term we shall be able to figure out what $\hat J^i$ should be. 
 
 Intuitively this is very similar to the aruguments used for the derivation of Wald entropy in case of higher derivative theories of gravity \cite{Wald},\cite{Iyer}. The gravity lagrangian there, has been replaced here by the expression of total entropy. If we vary a lagrangian, we generate a term which is directly proportional to the variation of the basic fields (identified as `equation of motion') along with a boundary term. In our case the `equation of motion' is replaced by the `equilibrium condition' and will always vanish by our starting assumption. The boundary term is analogous to our `$\hat J^i$'. Now if we substitute the variation of the basic fields (i.e. the functions appearing in the background metric and gauge field) with the Lie derivative along the original time-like Killing direction (which is no longer Killing once we are in an adiabatic situation) we shall get the equation \eqref{physeq}.

  $\hat J^0$ and $\hat J^i$ together will form a four-component current $\hat J^\mu$, constructed only from the background. Also by construction, $\hat J^0$ is of order ${\cal O}(\omega^0)$, $\hat J^i$ is of order ${\cal O}(\omega)$ and four-divergence  of $\hat J^\mu$ is non-zero only at order  ${\cal O}(\omega^2)$.

 We can see that $\hat J^\mu$  is constructed purely out of a very specific form of background metric and the gauge-field. Nowhere it involves the fluid variables. On the other hand the entropy current $J^\mu$ for a general fluid should be a function of the fluid variables and a background that is slowly varying but completely general otherwise. Therefore we could not identify $\hat J^\mu$  with fluid entropy current  $J^\mu$. But we know that $J^\mu$ must reduce to $\hat J^\mu$ upto order ${\cal O}(\omega)$ when we evaluate it on a background with very slow time dependence\footnote{If we want to evaluate $J^\mu$ on a background, we need to solve the fluid equations on  that background. But it is a difficult task. We shall use some trick to avoid this.  Instead of evaluating $J^\mu$ on the solution we shall evaluate it on a fixed profile for the fluid variables which is very close to the  solution but not exact. We shall try to argue that this would be enough for our purpose.}. In general we would expect this condition to fix the entropy current partially. By construction whenever the `equilibrium condition' is satisfied the divergence of this partially fixed entropy current will be of order ${\cal O}(\omega^2)$ if evaluated on a  background with a very slow time dependence.

Our next step would be to go away from the adiabatic regime. As mentioned in the previous section, at this stage we shall exactly compute the divergence of the entropy current we have determined from the adiabatic condition. By construction all the terms in this divergence will vanish whenever the system has a Killing time-like direction (i.e. in equilibrium). In addition to this, we also know that the divergence will vanish even in an adiabatic situation. Naively it implies that every term in this expression of divergence must contain at least two factors of the Killing equation (i.e. the Lie derivative of the background along the initial Killing direction) and/or its derivative. This is because in an adiabatic situation, the Killing equation which vanishes in perfect equilibrium, will evaluate to terms of order ${\cal O}(\omega)$. Schematically the divergence will be of the form
\begin{equation*}
\begin{split}
&D_\mu J^\mu|_\text{on time dependent background}
\sim \sum_{n\geq2}\sum_m(\partial)^m\bigg[ L_{\partial_t }(\text {Background})\bigg]^n
\end{split}
\end{equation*}
Where $L_{\partial_t }$ denotes the Lie derivative in the time direction or $\bigg[ L_{\partial_t }(\text {Background})\bigg]$ is essentially the Killing equations. The $m$ derivatives will be distributed among the $n$ factors of Killing equations in many different possible ways. 
\newline
 Once we have recasted the divergence in this form, it would be easy to figure out how we need to further modify the entropy current so that the final expression of the divergence is a positive definite form (see section(\ref{sec:explicit}) for details). We shall argue that such modification is always possible once the leading terms (in our case the divergence at second order in derivative expansion) are of appropriate signs.

 In summary our observations are the following.
 \begin{enumerate}
\item  If the background has very slow time dependence, it is possible to construct a current $\hat J^\mu$ (whose divergence is of order ${\cal O}(\partial_0^2)$), whenever the `equilibrium condition' is satisfied. 
\newline
$\hat J^\mu$ will be constructed in terms of a specific background with very slow time dependence.
\item Whenever $\hat J^\mu$ exists, we can construct an entropy current $J^\mu$ in terms of fluid variables whose divergence will be non negative on any slowly varying background provided some transport coefficients (ones that are unconstrained by the `equilibrium condition') satisfy some inequalities.
\item Only  the first order transport coefficients (in fact a subset of them\footnote{It turns out that in 4 dimensional anomalous fluid systems, some first order transport coefficients are determined by `equality type' relations, for example see \cite{son}}) have to satisfy some inequalities in order to ensure the `entropy condition'. It is not required to impose any `inequality type' constraint on the higher order transport terms.

 \end{enumerate}
 
 The main point here is that the existence of $\hat J^\mu$ require the `equilibrium condition' to be satisfied and once $\hat J^\mu$ exists, the `entropy condition' does not require any further `equality type' constraints.


\section{The set up}\label{outline}
In this section we shall explain the basic set-up and the notation that we shall use later in various sections. We shall mostly follow the notation and convention of \cite{equipart}.
\\

For the construction of $\hat J^\mu$ our set up is as follows.
\newline
The fluid lives in a (3+1) dimensional background that slowly varies in space and and also in time. However the time variation is slower than the space variation. In terms of equation what we mean is the following

\begin{equation}\label{equinot}
 \begin{split}
 &\text{Metric}: ~~ds^2 =G_{\mu\nu}dx^\mu dx^\nu = - e^{2\sigma}(dt + a_i dx^i)^2 + g_{ij}dx^i dx^j\\
 &\text{Gauge~field}:~ {\cal A} = A_0 dx^0 + {\cal A}_i dx^i = A_0 dt + (A_i + a_iA_0) dx^i\\
  &\text{where}\\
 &\sigma = \sigma(\vec x,\omega)e^{i\omega t},~a_i = a_i(\vec x,\omega)e^{i\omega t},~A_0 = A_0(\vec x,\omega)e^{i\omega t},~A_i = A_i(\vec x,\omega)e^{i\omega t}\\
 &\omega\ll 1\\
 &\text{Inverse length of the time circle at $\omega\rightarrow 0$} = T_0\\
 & \text{Holonomy around time circle at $\omega\rightarrow 0$} = A_0\\
 \end{split}
 \end{equation}
 Now we shall explain the notations\footnote{In equation \eqref{equinot} we have assumed that all functions in the background metric and the gauge field have a single frequency in time. This might seem too restrictive and it is actually not needed for the whole analysis as such. What we need is an independent  expansion in $\partial_t$ or $\partial_0$ operator, on top of the expansion in  terms of the space-derivatives. In \eqref{equinot} the derivative with respect to time is the smallest parameter. We shall use $\omega$ to count the number of time derivatives and we find it notationally simple if we have a single $\omega$ parameter for this purpose }.
 \newline
`$\bar\nabla_\mu$' denotes covariant derivative with respect to the full metric `$G_{\mu\nu}$' and `$\nabla_i$' denotes covariant derivative with respect to the spatial metric `$g_{ij}$'.
For the fluid variables we shall use $u^\mu,~T,~\mu$ to denote the 4-velocity, temperature and the chemical potential respectively. $u^\mu$ is normalized to $(-1)$. Instead of $\mu$ we shall often used $\nu$ as the independent variable, related to $\mu$ as $\nu = \frac{\mu}{T}$.   

Let us also fix some notations that we shall use later.
 \begin{equation}\label{notation}
 \hat u^\mu = e^{-\sigma}\{1,0,0,0\},~~\hat T = T_0 e^{-\sigma},~~\hat\nu = \frac{A_0}{T_0},~~\hat a_i = T_0 a_i
 \end{equation}
 In general if ${\cal B}(u^\mu, T,\nu)$ is some arbitrary function of fluid variables then by  $\hat{\cal B}$ we denote the same quantity evaluated on $\{\hat u^\mu,\hat T,\hat\nu\}$ and the background as given in equation \eqref{equinot}.
 $$\hat{\cal B}={\cal B}(\hat u^\mu,\hat T,\hat\nu)$$
 
 We should note that if we identify $\{u^\mu =\hat u^\mu,~ T=\hat T,~\nu = \hat\nu\}$, then it will solve the fluid equations only at ${\cal O}(\omega^0)$ but to all order in space derivatives (see  \cite{equipartamos} for a more detailed description on this fluid frame).
 
To construct $\hat J^\mu$,  we shall use the following decomposition for the stress tensor and the current in terms of the hatted quantities.
 \begin{equation}\label{stresscurrent2}
\begin{split}
T^{\mu\nu} &= (\hat E +\hat P) \hat u^\mu \hat u^\mu + \hat P G^{\mu\nu} + \tilde\pi^{\mu\nu}\\
C^\mu &= \hat Q \hat u^\mu +\tilde j^\mu\\
\end{split}
\end{equation} 
 In equation \eqref{stresscurrent2} $\hat E$, $\hat P$ and $\hat Q$ denote the expressions for the energy density, pressure and charge density respectively as derived from the zeroth order partition function. Outside strict equilibrium (strict $\omega\rightarrow 0$ limit) hatted quantities are not a solution to the fluid equations.
 Therefore $\tilde\pi^{\mu\nu}$ need not admit a local expansion in terms of the derivatives of $\{\hat u^\mu,\hat T,\hat\nu\}$ outside strict equilibrium.

 The partition function $Z$ is defined only in the limit $\omega\rightarrow 0$.  
 $$W=\log(Z )= \int \sqrt{g} L$$
  As usual $W$ is expressed in derivative expansion and it will have only space derivatives.
  $$L = L_{(0)} + L_{(1)}+L_{(2)}+\cdots$$
  Varying the partition function with respect to the metric and the gauge field we get the various components of the stress tensor  and the current evaluated on equilibrium \cite{equipart}, \cite{equipartamos}, \cite{equijustin}.
 \begin{equation}\label{stc}
\begin{split}
\lim_{\omega\rightarrow 0}\frac{\hat u^\mu \hat u^\nu T_{\mu\nu} }{\hat T^2} = \frac{1}{\sqrt{g}}\left[ \frac{\delta W}{\delta \hat T}\right],&~~
\lim_{\omega\rightarrow 0}\hat u^\mu  C_\mu = -\frac{1}{\sqrt{g}} \left[\frac{\delta W}{\delta \hat\nu}\right]\\
\lim_{\omega\rightarrow 0}\hat P^i_\mu  C^\mu = \frac{\hat T}{\sqrt{g}}\left[ \frac{\delta W}{\delta A_i}\right],&~~
\lim_{\omega\rightarrow 0}\frac{\hat P^i_\mu \hat u_\nu  T^{\mu\nu}}{\hat T^2} = \frac{1}{\sqrt{g}}\left[\frac{\delta W}{\delta (\hat a_i)}-\nu \frac{\delta W}{\delta A_i}\right] \\
\lim_{\omega\rightarrow 0}\frac{\hat P_{i\alpha}\hat  P_{j\beta}  T^{\alpha\beta} }{\hat T}  = -  \frac{2 }{\sqrt{g}}\left[ \frac{\delta W}{\delta g^{ij}}\right]&
\end{split}
\end{equation}
Here $\hat P^{\mu\nu} = \hat u^\mu \hat u^\nu + G^{\mu\nu}$.
In equation \eqref{stc} all quantities in the LHS are evaluated on equilibrium. \\

 Next we have to construct the full entropy current $J^\mu$ in a general background. By $D_\mu$ we shall denote the covariant derivative with respect to any arbitrary metric ${\cal G}_{\mu\nu}$.
 
For $J^\mu$, we shall use a different decomposition for the stress tensor and the current \cite{landau}. 
\begin{equation}\label{stresscurrent}
\begin{split}
T^{\mu\nu} &= E u^\mu u^\mu + P P^{\mu\nu} + \pi^{\mu\nu}\\
C^\mu &= Q u^\mu + j^\mu\\
&\text{where}~~P^{\mu\nu} = u^\mu u^\nu + {\cal G}^{\mu\nu}
\end{split}
\end{equation} 
$\pi^{\mu\nu}$ and $j^\mu$ will contain all the terms that have derivatives of the fluid variables. Therefore all the transport coefficients are contained in $\pi^{\mu\nu}$ and $j^\mu$. In equilibrium $E$, $P$ and $Q$ will evaluate to the energy density, pressure and the charge density of the system. They will be related to the equilibrium partition function by the usual laws of thermodynamics. Outside equilibrium their interpretation will depend on the choice of the fluid frame.
 
 We can see that $\tilde \pi^{\mu\nu}$ and $\tilde j^\mu$ (defined in equation \eqref{stresscurrent2}) are different from $\pi^{\mu\nu}$ and $j^\mu$ (defined in equation \eqref{stresscurrent}) if we go away from ($\omega\rightarrow 0$) limit. According to our notation
 \begin{equation*}
 \begin{split}
& \lim_{\omega\rightarrow 0}\tilde \pi^{\mu\nu} = \lim_{\omega\rightarrow 0}\pi^{\mu\nu} = \hat\pi^{\mu\nu},~~
 \lim_{\omega\rightarrow 0}\tilde j^\mu = \lim_{\omega\rightarrow 0}j^\mu = \hat j^\mu\\
\end{split}
\end{equation*}


\section{Entropy current as a function of background}\label{sec:general}
In this section we shall try to construct  the current $\hat J^\mu$ as described in section (\ref{sec:intro}).

 We shall first compute the total entropy in terms of a partition function expressed as a functional of the background metric and gauge field. Then we shall take the time derivative of the total entropy and using the `equilibrium condition' (i.e. equation \eqref{stc})  we shall explicitly show that it can be written as a pure boundary term as we have explained in section (\ref{sec:intro}). Finally from the expression of the total entropy and the boundary term we shall read off the current $\hat J^\mu$.
 
First we shall give a brief description of what we are going to do in the following subsections.
\begin{enumerate}
 \item The formula for total entropy in terms of the partition function is given as follows \cite{equipart}.
$$S_T = W +T_0 \frac{\partial W}{\partial T_0}$$
Here
\begin{equation}\label{notation2}
\begin{split}
&\text{partition function}=W = \int\sqrt{g}\left[L_{(0)} + L_{(1)}+ L_{(2)} +\cdots\right] = W_{(0)} + W_{(p)}\\
&\text{where} ~~W_{(0)}=\int\sqrt{g}L_{(0)} = \int\sqrt{g}\left( \frac{p(\hat T,\hat\nu)}{\hat T}\right)\\
& \text{with $p(\hat T,\hat\nu)$, some arbitrary function of its arguments} \\
&~~~~~~~~~W_{(p)} =\int\sqrt{g}\left[ L_{(1)}+ L_{(2)} +\cdots\right]=\int\sqrt{g}L_{(p)}
\end{split}
\end{equation}
\item We shall rewrite the total entropy in the following from
\begin{equation}\label{gen1}
 S_T = \int \sqrt{g} \left[s \hat u^0 - \frac{\hat u^\nu\tilde\pi^{0}_\nu}{\hat T} - \hat\nu \tilde j^0\right] +\sqrt{g}\left[ L_{(1)}+ L_{(2)} +\cdots\right]\hat u^0 + \sqrt{g}(\nabla_i \mathfrak K^i)
 \end{equation}
where 
$$s = \frac{\partial}{\partial T_0} \left (T_0 L_{(0)}\right)=L_{(0)} + T_0\left[\left(\frac{\partial\hat T}{\partial T_0}\right)\frac{\partial L_{(0)}}{\partial \hat T} +\left(\frac{\partial\hat \nu}{\partial T_0}\right)\frac{\partial L_{(0)}}{\partial \hat \nu}\right] = \frac{\partial p}{\partial\hat T}-\frac{\hat \nu}{\hat T}\frac{\partial p}{\partial\hat \nu}$$
 $\tilde\pi^{\mu\nu}$ and ${\tilde j^\mu}$ are defined in equation \eqref{stresscurrent2}.
  Here $\mathfrak K^i$ is some vector whose explicit form will depend on $W$, but we shall not need it for our construction (see the point (4) below). The content of equation \eqref{gen1} is just that the total entropy could be expressed as a sum of the first two terms in the RHS of equation \eqref{gen1} upto total derivatives.
\item Now we would like to see what constraints $\tilde\pi^{\mu\nu}$ and $\tilde j^\mu$ have to satisfy so that the rate of change of the total entropy with respect to time could be expressed purely as an influx of entropy through the boundary at leading order in $(\omega)$.

So we shall compute the time derivative of the total entropy and we shall do it term by term.

\item It is clear that $\partial_0$ of a total derivative piece (the last term in equation \eqref{gen1}) is trivially a boundary term and it does not require any constraints to be imposed on the stress tensor and the current. So we shall simply ignore it in the construction of the current.

\item Next we shall see that the time derivative of the first two terms  can be combined into a boundary term whenever the `equilibrium condition' is satisfied.

\item Finally from this boundary term we shall read off the space component of the current and we shall combine the time and the space component together in $\hat J^\mu$ in a covariant four-vector notation.
\end{enumerate}

 At this point we would like to point out that upto second order, relativistic hydrodynamics has already been studied in great detail through many different approaches (see \cite{RomatReview} and references therein). But here our main aim is to develop a general algorithm for the construction of the entropy current that is valid to all orders in derivative expansion. As we shall explain below, we need not truncate the derivative expansion to any  given order for our construction of $\hat J^\mu$.

\subsection{Total entropy in appropriate form}\label{sub:step2}
In this subsection we shall try to show equation \eqref{gen1}. This has been shown in \cite{equisuperfluid} in the context of superfluid with an anomalous current. For completeness we shall repeat the calculation here again but in a simpler situation of normal fluid without any anomaly.

As before 
\begin{equation}\label{before1}
W_{(0)} + T_0\left( \frac{\partial W_{(0)}}{\partial T_0}\right) = \int\sqrt{g}\left(\frac{\partial p}{\partial\hat T}-\frac{\hat \nu}{\hat T}\frac{\partial p}{\partial\hat \nu}\right) =\int\sqrt{g}~s
\end{equation}

Now we are going to compute the contribution of $W_{(p)}$ in total entropy. Ignoring the total derivative pieces we get the following.

\begin{equation}\label{superrep}
\begin{split}
&T_0 \frac{\partial W_{(p)}}{\partial T_0}\\
&= T_0\int_{\vec x}\left\{ \left( \frac{\delta W_{(p)}}{\delta \hat T(\vec x)}\right)\left(\frac{\partial \hat T(\vec x)}{\partial T_0}\right) +\left( \frac{\delta W_{(p)}}{\delta \hat a_i(\vec x)}\right)\left(\frac{\partial \hat a_i(\vec x)}{\partial T_0}\right)+\left( \frac{\delta W_{(p)}}{\delta \hat \nu(\vec x)}\right)\left(\frac{\partial \hat \nu(\vec x)}{\partial T_0}\right)
\right\}\\
&= T_0\int\sqrt{g}\left\{\left(\frac{\hat T}{T_0}\right)\left(\frac{\hat u^\mu \hat u^\nu\hat \pi_{\mu\nu} }{\hat T^2} \right)+ \frac{\hat a_i}{T_0}\left[\frac{\hat P^i_\mu \hat u^\nu \hat \pi^{\mu}_{\nu}}{\hat T^2} + \frac{\hat\nu}{\hat T}(\hat P^i_\mu \hat j^\mu)\right]+ \frac{\hat\nu}{T_0} (\hat u^\mu \hat j_\mu)\right\}\\
&= -\int \sqrt{G}\left[\frac{\hat u^\mu }{\hat T}\hat \pi^0_\mu + \hat\nu\hat j^0\right]
\end{split}
\end{equation}
In the third line we have used equation \eqref{stc} (ie. the `equilibrium condition') and in the fourth line we have used the explicit expressions for $\hat u^\mu,~\hat T,~\hat\nu$ and $\hat P^i_\mu$ in terms of the background (see equation \eqref{notation}).
$$\hat u^\mu = e^{-\sigma}\{1,0,0,0\},~~\hat T = T_0 e^{-\sigma},~~\hat\nu = \frac{A_0}{T_0},~~\hat P^{\mu\nu} = \hat u^\mu \hat u^\nu + G^{\mu\nu}$$
Using equation \eqref{before1} and \eqref{superrep} we can rewrite the total entropy in the desired form.
\begin{equation}\label{desfo}
\begin{split}
S_T &= \int \sqrt{G}\left[s\hat u^0 -\frac{\hat u^\mu}{\hat T} \hat \pi^0_\mu - \hat\nu\hat j^0\right] + W_{(p)}\\
&=\int \sqrt{G}\left[\left(s\hat u^0 -\frac{\hat u^\mu}{\hat T} \hat \pi^0_\mu - \hat\nu\hat j^0 \right)+ \hat u^0 L_{(p)}\right]\\
&=\int \sqrt{G}\left[\left(s\hat u^0 -\frac{\hat u^\mu}{\hat T} \tilde \pi^0_\mu - \hat\nu\tilde j^0\right)+ \hat u^0 L_{(p)}\right] + {\cal O}(\omega)
\end{split}
\end{equation}
In the last line we have used the decomposition of the stress tensor and current as given in equation \eqref{stresscurrent2} and used the fact that
$$\lim_{\omega\rightarrow0}\tilde\pi^{\mu\nu} = \hat\pi^{\mu\nu}\Rightarrow \tilde\pi^{\mu\nu} = \hat\pi^{\mu\nu} + {\cal O}(\omega),~~\lim_{\omega\rightarrow0}\tilde j^\mu = \hat j^\mu\Rightarrow \tilde j^\mu = \hat j^\mu + {\cal O}(\omega)$$

\subsection{Rate of entropy change at leading order in $(\omega)$}\label{sub:step3}
In this subsection we shall compute the derivative of the total entropy with respect to time at leading order in $\omega$. We shall see that once the `equilibrium condition' is satisfied, $(\partial_0 S_T)$ can be expressed as a pure boundary term.

As we have mentioned before, we shall compute this time derivative term by term. 

First we shall compute the time derivative of the first term in equation \eqref{desfo}. This can be processed using the conservation of the full stress tensor and current as written in equation \eqref{stresscurrent2}.
\begin{equation}\label{candiv}
\begin{split}
&\partial_0\int\left[\sqrt G\left(s\hat u^0 -\frac{\hat u^\mu}{\hat T} \tilde \pi^0_\mu - \hat\nu\tilde j^0\right)\right]\\
=& ~\int\partial_i\left[\sqrt{G} \left(\frac{\hat u^\nu}{\hat T}\tilde \pi^i_\nu + \hat\nu \tilde j^i\right)\right] -\int\sqrt{G}\left[ \tilde \pi^{\mu\nu} \bar\nabla_\mu \left(\frac{\hat u^\mu}{\hat T}\right)\right] +\int\sqrt{G}\left[\tilde j^\mu \left(\frac{\hat E_\mu}{\hat T} - \partial_\mu\hat\nu\right)\right]
\end{split}
\end{equation}
To derive equation \eqref{candiv} we have used the following thermodynamic relations between $\hat E,~\hat P,~s$ and $\hat T$ and $\hat\nu$.\footnote{These identities can be derived once we identify the stress tensor and the current at zero derivative order (the stress tensor and current of an ideal fluid) with variation of the zeroth order partition function with respect to the metric and the gauge field. In terms of equation this implies the following
$$-\frac{2T_0}{\sqrt{G}}\left(\frac{\delta W_{(0)}}{\delta G_{\mu\nu}}\right) =  \hat E \hat u^\mu \hat u^\nu + \hat P \hat P^{\mu\nu},~~~\frac{T_0}{\sqrt{G}}\left(\frac{\delta W_{(0)}}{\delta A_\mu}\right)= \hat Q \hat u^\mu$$} 
\begin{equation}\label{thermo2}
\begin{split}
d\Hat P &= s d\hat T + \hat Q d\hat\mu\\
\hat E + \hat P &= \hat T s + \hat\mu \hat Q,~~
\text{where}~\hat \mu = \hat T\hat \nu
\end{split}
\end{equation}
We shall rewrite the last two terms in the RHS of equation \eqref{candiv} in the following way.
\begin{equation}\label{lasttwo}
\begin{split}
&\int\sqrt{G}\left[- \tilde \pi^{\mu\nu} \bar\nabla_\mu \left(\frac{\hat u^\mu}{\hat T}\right) +\tilde j^\mu \left(\frac{\hat E_\mu}{\hat T} - \partial_\mu\hat\nu\right)\right]\\
=&-\int\sqrt{G}\bigg\{\frac{\hat\Theta}{3T} \left(\hat P_{ij}\hat \pi^{ij} \right)+\left(\frac{\hat u^\mu \hat u^\nu \hat \pi_{\mu\nu} }{\hat T^2}\right)(\hat u.\partial \hat T) - \left(\hat u^\mu \hat j_\mu\right)(\hat u.\partial \hat \nu)\\
&~~~~~~~~~~~~~- \left(\frac{\hat P^i_\mu \hat u_\nu \hat \pi^{\mu\nu}}{\hat T^2}\right)\hat h_i -\left(\hat P^i_\mu \hat j^\mu\right)\hat v_i +\left(\frac{\hat P^i_{\alpha}\hat  P^j_{\beta} \hat \pi^{\alpha\beta} }{\hat T}\right)\hat\sigma_{ij}\bigg\} + {\cal O}(\omega^2)
\end{split}
\end{equation}
where
 \begin{equation}\label{notprofhat}
 \begin{split}
 &\hat P^{\mu\nu} = G^{\mu\nu} + \hat u^\mu \hat u^\nu\\
 &\hat \Theta = \bar \nabla_\mu \hat u^\mu\\
 &\hat h_\mu = (\hat u^\nu\bar\nabla_\nu) \hat u^\mu +\hat P_\mu ^\alpha\frac{\bar\nabla_\alpha \hat T}{\hat T} \\
 &\hat v_\mu = \frac{\hat E_\mu }{\hat T}
 - \hat P_\mu^\alpha\bar\nabla_\alpha \hat \nu = \frac{\hat{\cal F}_{\mu\nu}\hat u^\nu }{\hat T}- \hat P_\mu ^\alpha\bar\nabla_\alpha \hat \nu,~~\hat{\cal F}_{\mu\nu} = \bar\nabla_\mu {\cal A}_\nu- \bar\nabla_\nu {\cal A}_\mu \\
 &\hat \sigma_{\mu\nu} = \hat P_\mu^{\alpha} \hat P_\nu^\beta\left[\frac{\bar\nabla_\alpha \hat u_\beta +\bar\nabla_\beta \hat u_\alpha }{2}- \frac{\hat \Theta}{3} G_{\alpha\beta}\right]
 \end{split}
 \end{equation}
 In equation \eqref{lasttwo} we have again used the $\omega$ expansion  for $\tilde \pi^{\mu\nu}$ and $\tilde j^\mu$.
 $$\tilde\pi^{\mu\nu} = \hat\pi^{\mu\nu} + {\cal O}(\omega),~~\tilde j^\mu = \hat j^\mu + {\cal O}(\omega)$$
 Combining equations \eqref{candiv} and \eqref{lasttwo} we finally get the time derivative for the first term in total entropy as given in equation \eqref{desfo}.
 \begin{equation}\label{finalcandiv}
 \begin{split}
 &\partial_0\int\left[\sqrt G\left(s\hat u^0 -\frac{\hat u^\mu}{\hat T} \tilde \pi^0_\mu - \hat\nu\tilde j^0\right)\right]\\
=&~-\int\sqrt{G}\bigg\{\frac{\hat\Theta}{3T} \left(\hat P_{ij}\hat \pi^{ij} \right)+\left(\frac{\hat u^\mu \hat u^\nu \hat \pi_{\mu\nu} }{\hat T^2}\right)(\hat u.\partial \hat T) - \left(\hat u^\mu \hat j_\mu\right)(\hat u.\partial \hat \nu)\\
&~~~~~~~~~~~~~- \left(\frac{\hat P^i_\mu \hat u_\nu \hat \pi^{\mu\nu}}{\hat T}\right)\hat h_i -\left(\hat P^i_\mu \hat j^\mu\right)\hat v_i +\left(\frac{\hat P^i_{\alpha}\hat  P^j_{\beta} \hat \pi^{\alpha\beta} }{\hat T}\right)\hat\sigma_{ij}\bigg\} \\
& ~+\int\partial_i\left[\sqrt{G} \left(\frac{\hat u^\nu}{\hat T}\tilde \pi^i_\nu + \hat\nu \tilde j^i\right)\right]+ {\cal O}(\omega^2)
 \end{split}
 \end{equation}
See appendix (\ref{derivation}) for a detailed derivation of equations \eqref{candiv} and \eqref{lasttwo}.

Now we are going to compute the time derivative of the second term in the total entropy.

A schematic form of $\partial_0 W_{(p)}$ will be the following.
 \begin{equation}\label{kichhuna2}
\begin{split}
\partial_0 \left[\sqrt{g} L_{(p)}\right] =\sum_{n=0}^{N_\Phi} B^{\Phi}_{j_1,j_1,\cdots,j_n}[\partial_{j_1}\partial_{j_2}\cdots\partial_{j_n}\partial_0 \Phi]
\end{split}
\end{equation}
Where $\Phi$ collectively denotes  $\{\hat T,~ \hat \nu,~\hat a_i,~A_i,~g^{ij}\}$ and $B^{\Phi}_{j_1,j_1,\cdots,j_n} = \frac{\partial\left[\sqrt{g} L_{(p)}\right]}{\partial [\partial_{j_1}\partial_{j_2}\cdots\partial_{j_n}\Phi]}$.

Now any term with a form as described in equation \eqref{kichhuna2} can be re-written as a sum of two terms, one being proportional to $\partial_0\Phi$ and the other is a total derivative.

\begin{equation}\label{kichhuna3}
\begin{split}
&\int B^{\Phi}_{j_1,j_1,\cdots,j_n}[\partial_{j_1}\partial_{j_2}\cdots\partial_{j_n}\partial_0 \Phi]\\
 \sim &(-1)^n\int[\partial_{j_1}\partial_{j_2}\cdots\partial_{j_n}B^{\Phi}_{j_1,j_1,\cdots,j_n}](\partial_0\Phi) +\int \partial_i (\sqrt{g}U^i)\\
 \sim &\left(\frac{\delta W_{(p)}}{\delta \Phi}\right)\partial_0\Phi +\int \partial_i (\sqrt{g}U^i)\\
 \end{split}
 \end{equation}
 In the last line of \eqref{kichhuna3} the first term is the functional derivative of  $W_{(p)}$ in the bulk. Here the functional derivative is taken  assuming that the infinitesimal change in the field $(\delta\Phi\sim\partial_0\Phi)$ vanishes in the boundary (ie. ignoring the total derivative pieces or boundary terms). The second term gives the boundary current $U^i$.
  It is equal to the change of the partition function exactly on the boundary due to the change in the background field proportional to $\delta \Phi\sim\partial_0\Phi$. In other words $U^i$ contains the terms that we have ignored while computing the first term in equation \eqref{kichhuna3}.
  
 From the above argument it is clear that $U^i$ will have a unique expression in terms of the background upto terms whose space divergence identically vanish. And it will generically have a factor of the form ($\partial_0 \Phi$) and hence is of order ${\cal O}(\omega)$.

Now using equation \eqref{stc} we can relate the first term in equation \eqref{kichhuna3} to the equilibrium values of the stress tensor and the current i.e. $(\hat\pi^{\mu\nu})$ and $(\hat j^\mu)$. Substituting we finally get the following expressions for the time derivative of the second term in the total entropy.
\begin{equation}\label{kichhu1}
\begin{split}
&\partial_0\left(\int\sqrt{g} L_{(p)}\right)-\int \partial_i (\sqrt{g}U^i)\\
=&\int\left[  \left(\frac{\delta }{\delta g^{ij}}[\sqrt{g}L_{(p)}]\right)(\partial_0 g^{ij}) + \sqrt{g}\left(\frac{\delta L_{(p)}}{\delta \hat T}\right)(\partial_0 \hat T) +\sqrt{g} \left(\frac{\delta L_{(p)}}{\delta \hat\nu}\right)(\partial_0 \hat \nu)\right]\\
&+\int\sqrt{g}\left[ \left(\frac{\delta L_{(p)}}{\delta \hat a_i}\right)(\partial_0 \hat a_i) +\left(\frac{\delta L_{(p)}}{\delta A_i}\right)(\partial_0 A_i)\right]\\
=&\int\sqrt{G}\bigg\{\frac{\hat\Theta}{3T} \left(\hat P_{ij}\hat \pi^{ij} \right)+\left(\frac{\hat u^\mu \hat u^\nu \hat \pi_{\mu\nu} }{\hat T^2}\right)(\hat u.\partial \hat T) - \left(\hat u^\mu \hat j_\mu\right)(\hat u.\partial \hat \nu)\\
&~~~~~~~~~~~~~- \left(\frac{\hat P^i_\mu \hat u_\nu \hat \pi^{\mu\nu}}{\hat T}\right)\hat h_i -\left(\hat P^i_\mu \hat j^\mu\right)\hat v_i +\left(\frac{\hat P^i_{\alpha}\hat  P^j_{\beta} \hat \pi^{\alpha\beta} }{\hat T}\right)\hat\sigma_{ij}\bigg\}
\end{split}
\end{equation}
In the last line of equation \eqref{kichhu1} we have used the explicit expressions for all the hatted quantities. See appendix (\ref{derivation}) for the derivation of equation \eqref{explicit}.
 \begin{equation}\label{explicit}
 \begin{split}
 \partial_0 g^{ij}&= -2 e^{\sigma}\left(\hat\sigma^{ij} + \frac{g^{ij}}{3} \hat \Theta\right),~~\partial_0\sqrt{g} = e^{\sigma}\sqrt{g}\hat\Theta\\
 \partial_0 A_j &= -g_{ji}\left(\hat v^i - \nu\hat  h^i\right)\hat T e^{\sigma},
 ~~\partial_0\hat a_j = - g_{ji}\hat T e^{\sigma}\hat h^i\\
 \partial_0\hat T &= e^{\sigma}(\hat u.\partial\hat T) ,
 ~~\partial_0\hat \nu = e^{\sigma}(\hat u.\partial\hat \nu) 
 \end{split}
 \end{equation}
 
Now combining equations \eqref{kichhu1}, \eqref{candiv} and \eqref{lasttwo} and using the fact that 
$$\hat P_{ij} = g_{ij},~~\hat P^i_\mu = \delta^i_\mu$$ 
we get the time derivative of the total entropy and we see that the bulk terms cancel upto order ${\cal O}(\omega)$ . Therefore the rate of total entropy change in the bulk can be expressed purely as a boundary term.
\begin{equation}\label{finalent}
\begin{split}
\partial_0 S_T&=\partial_0\int_{space}\left[\sqrt{G}\left(s\hat u^0 -\frac{\hat u^\mu \tilde \pi^0_\mu }{\hat T}- \hat\nu\tilde j^0 + L_{(p)} \hat u^0\right) \right]+ {\cal O}(\omega^2)\\
 &= \int_{space}\partial_i\left[\sqrt{G}\left(\frac{\hat u_\nu\tilde\pi^{i\nu}}{\hat T} + \hat\nu\tilde j^i\right)\right]+\int\partial_i\left[\sqrt{G} e^{-\sigma}U^i\right]+ {\cal O}(\omega^2)\\
\end{split}
\end{equation}
To arrive at equation \eqref{finalent} we have used equation \eqref{stc} at several places which is equivalent to the `equilibrium condition'. Secondly we have assumed that $\tilde\pi^{\mu\nu}$ and $\tilde j^\mu$ have an analytic expression in $(\omega)$ for the states that are slowly varying in time. In terms of equation this means the following
\begin{equation}\label{impassump}
\begin{split}
&\lim_{\omega\rightarrow 0}\tilde\pi^{\mu\nu} = \hat\pi^{\mu\nu},~~\lim_{\omega\rightarrow 0}\tilde j^{\mu} = \hat j^{\mu}\\
\Rightarrow &\tilde\pi^{\mu\nu} = \hat\pi^{\mu\nu} + {\cal O}(\omega),~~\tilde j^{\mu} = \hat j^{\mu}+ {\cal O}(\omega)\\
\end{split}
\end{equation}
But we can see that to derive equation \eqref{finalent}, $\tilde\pi^{\mu\nu}$ and $\tilde j^\mu$ need not admit a derivative expansion in terms of the derivatives of $\hat u^\mu$, $\hat T$ and $\hat \nu$.

Finally we can read off the time and the space component of $\hat J^\mu$ from equation \eqref{finalent}.
\begin{equation}\label{gencur}
\begin{split}
\hat J^0 &= \left(s\hat u^0 -\frac{\hat u^\mu \tilde \pi^0_\mu }{\hat T}- \hat\nu\tilde j^0 + L_{(p)} \hat u^0\right)\\
\hat J^i &=-\left(\frac{\hat u^\nu\tilde\pi^{i}_{\nu}}{\hat T} + \hat\nu\tilde j^i\right)-e^{-\sigma}U^i
\end{split}
\end{equation}
For convenience here we shall repeat the definition of $L_{(p)}$ and $U^i$ in words.
$L{(p)}$ is a scalar function of the background metric and the gauge field where each term contains at least one space derivative. Integration of $L_{(p)}$ over space gives the generating function for the stress tensor and the currents at one or higher order in derivative expansion. We denote the generating function as $W_{(p)}$.
$$W_{(p)} = \int \sqrt{g} L_{(p)}$$ 
As the background metric and the gauge field vary slowly with time, $W_{(p)}$ also varies. Inside the bulk of the space, the variation of $W_{(p)}$  is captured by the equilibrium values of the stress tensor and the current. And along the boundary of the space the variation could naturally be expressed as a surface integration of some  vector. This is the vector which we denote as $U^i$. It is defined only upto some `exact' vectors whose integration over the boundary identically vanish. This `exact' piece could be non-vanishing even in equilibrium. But the rest of the $U^i$ must be proportional to the time derivative of the background.

Now we shall try to combine  $\hat J^0$ and $\hat J^i$ in a covariant four-vector $\hat J^\mu$. The part that involves $s$, $\tilde \pi^{\mu\nu}$ and $\tilde j^\mu$ can be easily covariantized. Let us denote this part as $\hat J^\mu_{can}$.
\begin{equation}\label{curcan}
\begin{split}
\hat J^\mu &= \hat J^\mu_{can} + \hat S^\mu\\
\text{where}~~\hat J^\mu_{can} &= s\hat u^\mu - \frac{\hat u^\nu\tilde\pi^{\mu}_{\nu}}{\hat T} + \hat\nu\tilde j^\mu\\
\end{split}
\end{equation}
According to equation \eqref{curcan}, $\hat S^\mu$ is the covariantized version of the part involving $L_{(p)}$ and $U^i$.
At this point it would be interesting to note that the construction of $\hat S^\mu$ is very much like the construction of  Noether current corresponding to the diffeormorphism symmetry of $W_{(p)}$ along `time' direction. Such  Noether current is the starting point for computing Wald entropy of higher derivative gravity theory \cite{Wald},\cite{Iyer}. In \cite{Wald} or \cite{Iyer} this current is conserved upto equations of motion i.e. the linear variation of the gravity `action' with respect to the metric and other dynamical fields. Of course, in our case the metric is not dynamical and we cannot set the bulk term  $\left[\left(\frac{\delta W_{(p)}}{\delta \Phi}\right)\partial_0\Phi\right] $ (which is analogous to the `equations of motion' in \cite{Wald}) to 
zero. However here the `equations of motion' is replaced by the `equilibrium condition'. We have seen that once the `equilibrium condition' is satisfied, the current is conserved upto the leading order in adiabatic expansion.

It is not possible to write a general covariant expression of $\hat S^\mu$  as we have done it for $\hat J^\mu_{can}$. We can do it only in a case by case basis\footnote{It might be possible that the part of the entropy current involving $L_{(p)}$ and $U^i$ could be covariantized only after we add some further `exact' terms to $U^i$ whose integral over the boundary vanish. This could be the case for parity odd situations. 
\newline
For example, if $L_{(p)}$ is of the form
$$L_{(p)} \sim \epsilon^{ijk} a_i \partial_j A_k$$
then the only way to covariantize it would be to add a new term to $U^i$ of the form
$$U^i_{new}\sim \epsilon^{ijk} \partial_j A_k$$
Integration of this $U^i$ over the boundary will vanish and therefore it will not affect the total entropy influx or outflux through the boundary.}.
However for our purpose we shall not need any explicit expression for $\hat S^\mu$.


\section{Entropy current in terms of fluid variables}\label{sec:fluident}
The current $\hat J^\mu$ that we have constructed so far is a function of the background geometry. But the entropy current $J^\mu$ is generally  a function of the fluid variables like the velocity, temperature or the chemical potential. In this section we shall see how we can extend $\hat J^\mu$ to the full entropy current $J^\mu$. 

 According to the `entropy condition', $J^\mu$ should satisfy the following properties.
\begin{enumerate}
\item In a time independent situation the integration of $J^0$ over any spatial slice must give the total entropy of the system.
\item Divergence of $J^\mu$ should be non negative for any fluid flow consistent with the equations of motion.

As a consequence, for a very slow time dependence, the divergence of $J^\mu$ must vanish till order ${\cal O}(\omega)$. 
\end{enumerate}
Generically $J^\mu$ is a function of the fluid variables and if we want to evaluate it or its divergence on any time dependent background we need to know the solution of the fluid equations on that background.

It is difficult to solve the fluid equations exactly. But we need the solution only upto  order ${\cal O}(\omega)$. Here we shall assume that in our case there exists at least one time dependent but approximate solution where the fluid variables simply follow the equilibrium (slowly shifting with time). In terms of equations we mean the following.

 If $u^\mu$, $T$ and $\nu$ are the solutions for fluid velocity, temperature and the chemical potential respectively then they can be written as
\begin{equation}\label{assump}
\begin{split}
u^\mu = \hat u^\mu + {\cal O}(\omega),~~T = \hat T + {\cal O}(\omega),~~\nu = \hat \nu + {\cal O}(\omega)
\end{split}
\end{equation}
Through equation \eqref{assump} we also partially fix the fluid frame. This equation implies that the fluid frame that we  are  choosing is the one that reduces to $\{\hat u^\mu,\hat T, \hat\nu\}$ in equilibrium (see \cite{equipartamos},\cite{equiloga},\cite{loga1},\cite{loga2} for the use of this fluid-frame in case of parity odd terms in anomalous fluids). We shall have some more discussion about the existence of such solution in  appendix (\ref{app:solution}).

Now we shall first write a prescription for how to construct $J^\mu$ with the help of $\hat J^\mu$ already determined. Next we shall see how this way of constructing $J^\mu$  is consistent with the `entropy condition'.
\subsection{The prescription to go from $\hat J^\mu$ to $J^\mu$}
\begin{enumerate}
\item We shall decompose $J^\mu$ in two parts as we have done for $\hat J^\mu$ in equation \eqref{curcan}.
$$J^\mu = J^\mu_{can} + S^\mu$$
\item Next we demand that when evaluated on $\{\hat u^\mu,\hat T,\hat \nu\}$,  $J^\mu_{can}$  and $S^\mu$ should reduce to $\hat J^\mu_{can}$ and $\hat S^\mu$ respectively.
\item The obvious choice for $J^\mu_{can}$ is the following.
$$J^\mu_{can} = s u^\mu -\frac{u_\nu\pi^{\mu\nu}}{T} - \nu j^\mu$$
Here we have used equation \eqref{stresscurrent} for the decomposition of the stress tensor and the current in ideal and derivative part.
\item For $S^\mu$ we shall write the most general covariant vector expression possible at a given derivative order. The number of independent terms would be equal to the number of on-shell independent scalars or vectors constructed out of the fluid variables and background at that particular order in derivative expansion.
\item Next we shall evaluate it on $\{\hat u^\mu,\hat T,\hat \nu\}$ and compare the result with $\hat S^\mu$  upto order ${\cal O}(\omega)$. This will fix several unfixed coefficients in $S^\mu$ in terms of the coefficients in partition function.
\item The `entropy condition' will further constrain the remaining coefficients. However these new constraints will not depend on the partition function.
\end{enumerate}

\subsection{The divergence of $J^\mu$ and the `entropy condition'}
In this subsection we shall partially compute the divergence of $J^\mu$ using our knowledge about the divergence of $\hat J^\mu$.

First we shall compute the divergence for the canonical part. Using the conservation of stress tensor and the current we can calculate the 
divergence exactly\footnote{Equation \eqref{ent:divcan} has been derived and used in several cases before, see for instance \cite{son}, \cite{superfluident}}.
\begin{equation}\label{ent:divcan}
 \begin{split}
 J^\mu_{can} &= s u^\mu - \frac{u_\nu\pi^{\mu\nu}}{T}-\nu j^\mu\\
 D_\mu J^\mu_{can} &= (j^\mu u_\mu)(u.\partial\nu) -\left(\frac{u_\mu u_\nu \pi^{\mu\nu}}{T^2} \right)(u.\partial T) -\left(\frac{P_{\mu\nu}\pi^{\mu\nu}}{3T} \right)\Theta\\
 &+v_\mu j^\mu + \left(\frac{u_\nu \pi^{\mu\nu}}{T}\right)h_\mu -\left(\frac{\pi^{\mu\nu}}{T}\right)\sigma_{\mu\nu}
 \end{split}
 \end{equation}
 Here
 \begin{equation}\label{ent:notprof}
 \begin{split}
 &P^{\mu\nu} = {\cal G}^{\mu\nu} + u^\mu u^\nu\\
 &\Theta = D_\mu u^\mu\\
 &h_\mu = (u^\nu D_\nu) u^\mu +P_\mu ^\alpha\left(\frac{ D_\alpha T }{T}\right)\\
 &v_\mu = \frac{{\cal F}_{\mu\nu} u^\nu }{T}- P_\mu ^\alpha D_\alpha \nu\\
 &\sigma_{\mu\nu} = P_\mu^{\alpha} P_\nu^\beta\left[\frac{D_\alpha u_\beta +D_\beta u_\alpha }{2}- \frac{\Theta}{3} {\cal G}_{\alpha\beta}\right]
 \end{split}
 \end{equation}
 Here $D_\mu$ denotes the covariant derivative with respect to the general background ${\cal G}_{\mu\nu}$.
 
By explicit computation we can see that $\Theta$, $h_\mu$, $v^\mu$ and $\sigma_{\mu\nu}$ vanish in strict equilibrium. Hence if we evaluate these quantities on $\{\hat u^\mu, \hat T, \hat \nu\}$ and in the background as given in equation \eqref{equinot}, they will be of order ${\cal O}(\omega)$.
 
 Now we have to compute the divergence of $S^\mu$.  First we should note that using the equation\eqref{kichhu1} we could explicitly determine  what the divergence of $\hat S^\mu$ should be.
 \begin{equation}\label{ent:divhats}
 \begin{split}
 \bar\nabla_\mu \hat S^\mu &=\bigg\{\frac{\hat\Theta}{3T} \left(\hat P_{ij}\hat \pi^{ij} \right)+\left(\frac{\hat u^\mu \hat u^\nu \hat \pi_{\mu\nu} }{\hat T^2}\right)(\hat u.\partial \hat T) - \left(\hat u^\mu \hat j_\mu\right)(\hat u.\partial \hat \nu)\\
&~~~~~~~~~~~~~- \left(\frac{\hat P^i_\mu \hat u_\nu \hat \pi^{\mu\nu}}{\hat T}\right)\hat h_i -\left(\hat P^i_\mu \hat j^\mu\right)\hat v_i +\left(\frac{\hat P^i_{\alpha}\hat  P^j_{\beta} \hat \pi^{\alpha\beta} }{\hat T}\right)\hat\sigma_{ij}\bigg\}\\
&=\bigg\{\frac{\hat\Theta}{3T} \left(\hat P_{\mu\nu}\hat \pi^{\mu\nu} \right)+\left(\frac{\hat u^\mu \hat u^\nu \hat \pi_{\mu\nu} }{\hat T^2}\right)(\hat u.\partial \hat T) - \left(\hat u^\mu \hat j_\mu\right)(\hat u.\partial \hat \nu)\\
&~~~~~~~~~~~~~- \left(\frac{\hat P^\beta_\mu \hat u_\nu \hat \pi^{\mu\nu}}{\hat T^2}\right)\hat h_\beta -\left(\hat P^\beta_\mu \hat j^\mu\right)\hat v_\beta +\left(\frac{\hat P^{\nu}_{\alpha}\hat  P^{\mu}_{\beta} \hat \pi^{\alpha\beta} }{\hat T}\right)\hat\sigma_{\mu\nu}\bigg\}
 \end{split}
 \end{equation}
 where
 \begin{equation}\label{ent:notprofhat}
 \begin{split}
 &\hat P^{\mu\nu} = G^{\mu\nu} + \hat u^\mu \hat u^\nu\\
 &\hat \Theta = \bar \nabla_\mu \hat u^\mu\\
 &\hat h_\mu = (\hat u^\nu\bar\nabla_\nu) \hat u^\mu +\hat P_\mu ^\alpha\left(\frac{\bar\nabla_\alpha \hat T}{\hat T}\right) \\
 &\hat v_\mu = \frac{\hat E_\mu }{\hat T}- \hat P_\mu ^\alpha\bar\nabla_\alpha \hat \nu\\
 &\hat \sigma_{\mu\nu} = \hat P_\mu^{\alpha} \hat P_\nu^\beta\left[\frac{\bar\nabla_\alpha \hat u_\beta +\bar\nabla_\beta \hat u_\alpha }{2}- \frac{\hat \Theta}{3} G_{\alpha\beta}\right]
 \end{split}
 \end{equation}
 In the second line of equation \eqref{ent:divhats} we have used the fact that\footnote{ This is true because by definition $\hat u^\mu \hat h_\mu=\hat u^\mu \hat v_\mu =0$ and $\hat u^\mu\hat\sigma_{\mu\nu} = \hat u^\mu \hat P_{\mu\nu} =0$} 
 $$\hat h_0 = \hat v_0 =0,~~~\hat \sigma_{00} = \hat \sigma_{0i} =\hat P_{00} = \hat P_{0i}=0$$
Clearly in equation\eqref{ent:divhats} we do not need $\{\hat u^\mu, \hat T, \hat \nu\}$ to satisfy any equations of motion; this is just an identity relating the derivatives of the background functions.

Suppose we also compute the divergence of $S^\mu$ without using any equations of motion, i.e. we treat each space-time derivative of the fluid variables as locally independent piece of data\footnote{In other words, the computation that we have in mind is just a rewriting of $D_\mu S^\mu$ in terms of our chosen basis of off-shell independent scalars at that particular order in derivative expansion. It is an identity and therefore true for any profile of fluid variables even if they do not satisfy the conservation equations.}. Naively the final expression of the divergence would contain all the off-shell independent scalars that could be constructed out of the fluid variables and the metric and gauge field upto the derivative order we are interested in.
However we also know that $S^\mu$ reduces to $\hat S^\mu$ when evaluated on $\{\hat u^\mu, \hat T, \hat \nu\}$ upto some appropriate order in $\omega$.
It follows that $D_\mu S^\mu$ must evaluate to equation \eqref{ent:divhats} when evaluated on $\{\hat u^\mu, \hat T, \hat \nu\}$ upto order ${\cal O}(\omega^2)$. Therefore  we expect that the divergence of $S^\mu$  could be rewritten in the following form (which would just be an identity relating the derivatives of the fluid variables)
 \begin{equation}\label{ent:divs}
 \begin{split}
 D_\mu S^\mu &= S_c(u.\partial\nu)+ S_T(u.\partial T) 
 +A^{\mu\nu}\sigma_{\mu\nu}
 +S_\pi\Theta
 +  K^\mu v_\mu + H^\mu h_\mu\\
 &  + \text{terms that are of ${\cal O}(\omega^2)$ on $\{\hat u^\mu, \hat T, \hat \nu\,\}$ and $\{ G_{\mu\nu}, {\cal A}_\mu\}$}\\
 \end{split}
 \end{equation}

 Where $S_c, ~S_T,~S_\pi,~K^\mu,~H^\mu$ and $A_{\mu\nu}$  are some scalar, vector and tensor functions of the fluid variables satisfying
 \begin{equation}\label{eqcon}
 \begin{split}
 &\hat S_c = -\hat j^\mu \hat u_\mu ,~~ \hat S_T  = \frac{\hat u^\mu \hat u^\nu \hat \pi_{\mu\nu}}{\hat T^2},~~\hat S_\pi=\frac{\hat P_{ij}\hat \pi^{ij}}{3\hat T} \\
&\hat K^i=-\hat j^i,~~\hat H^i =-\frac{\hat u^\mu \hat \pi_{\mu}^{ i}}{\hat T},~~\hat A^{ij}=\frac{\hat \pi^{ij}}{T}\\
 \end{split}
 \end{equation}
 As usual by $\hat S_c, ~\hat S_T,~\hat S_\pi,~\hat K^\mu,~\hat H^\mu$ and $\hat A_{\mu\nu}$ we denote the ${\cal O}(\omega^0)$ piece of $S_c, ~S_T,~S_\pi,~K^\mu,~H^\mu$ and $A_{\mu\nu}$ respectively, when evaluated on $\{\hat u^\mu, \hat T, \hat \nu\}$ in the background as given in equation \eqref{equinot}.
 
The equation \eqref{ent:divs} should be true for any profile of the velocity ($u^\mu$), temperature ($T$) and the chemical potential ($\nu$) and therefore in particular the ones that satisfy the equations of motion (whereas the equation \eqref{ent:divcan} is valid only if the fluid
  variables satisfy the equations of motion). Also from our assumption about the solution as given in equation \eqref{assump} it follows that for any function of fluid variables, the order of $\omega$ cannot be lowered by substituting $\{\hat u^\mu, \hat T, \hat \nu\}$ in place of $\{ u^\mu,  T,  \nu\}$. In other words if a term, when evaluated on $\{\hat u^\mu, \hat T, \hat \nu\}$, is of the order ${\cal O}(\omega^2)$ then it will remain of order ${\cal O}(\omega^2)$ or higher when evaluated on actual fluid solution. Hence we can rewrite equation \eqref{ent:divs} as
  \begin{equation}\label{ent:divsf}
 \begin{split}
D_\mu S^\mu &= S_c(u.\partial\nu)+ S_T(u.\partial T) 
 +A^{\mu\nu}\sigma_{\mu\nu}
 +S_\pi\Theta
 +  K^\mu v_\mu + H^\mu h_\mu\\
 & + \text{terms that are of ${\cal O}(\omega^2)$ on $\{ G_{\mu\nu}, {\cal A}_\mu\}$}
 \end{split}
 \end{equation}
 Where the first six terms are of order ${\cal O}(\omega)$.
 
We can combine equations \eqref{ent:divcan} and \eqref{ent:divsf} to compute the divergence of the full entropy current $J^\mu$, The following seems to be true by construction.
 \begin{enumerate}
 \item $D_\mu J^\mu$ is of order ${\cal O}(\omega^2)$ if we evaluate it on the solutions of fluid equations in a  background as given in equation \eqref{equinot}.
 \item All those terms in the stress tensor and current that are non-zero in equilibrium are not there in the final expression of the divergence.
  \end{enumerate}

  Next we would like to show that once the equation \eqref{ent:divsf} is true, we can always find a a choice of coefficients for the rest of the unfixed terms in $S^\mu$ such that the `entropy condition' is satisfied. 
  This choice will depend on those parts of the stress tensor and current that vanish in equilibrium, but its existence will not require any further `equality type' constraints among the transport coefficients. In section (\ref{sec:example})we shall explicitly see how it works for a simple example.  
 
  We shall mainly follow the line of arguments presented in \cite{secondorder}. We have  to express the final divergence as a sum of squares.  
  However now we have to do the analysis for an arbitrary background and in this case it is very difficult to say anything general about the solutions of the fluid equations.  So we have to argue with the minimum use of the equations of motion.  
   
First we note the following features which could lead to a rewriting of the divergence as required. 
  \begin{enumerate}
  \item Transport coefficients enter the expression of divergence only through the divergence of the canonical part.
  \item Only those parts  of the stress tensor and the current that vanish in equilibrium, will enter the expression of divergence.

   Let us call all these terms collectively as `dissipative terms' and corresponding transport coefficients as `dissipative coefficients'.

\item   Other than $(u.\partial T)$, $(u.\partial\nu)$,  $\sigma_{\mu\nu}$, $\Theta$ , $h_\mu$ and $v^\mu$ all other dissipative terms will appear linearly in  the expression of the divergence of the canonical part ( always multiplied by at least one factor of $(u.\partial T)$, $(u.\partial\nu)$,  $\sigma_{\mu\nu}$, $\Theta$ , $h_\mu$ or $v^\mu$)\footnote{Since we are not using any equations of motion $(u.\partial T)$, $(u.\partial\nu)$ and $\Theta$ will always be treated as independent. The same is true for $h_\mu$ and $v^\mu$.}.

 \end{enumerate}

 Now consider a fluid profile where locally only one such dissipative term is non zero. Since this term is dissipative, it must produce entropy and also according to the assumption of fluid dynamics and  the `entropy condition', this entropy should be produced locally. Hence apart from the linear pieces, the divergence must contain quadratic or higher order pieces of the dissipative terms as well.

 Because the quadratic and higher order pieces of the dissipative terms are present, it should be possible to rewrite the divergence as sum of squares with each square term multiplied by some positive definite coefficient.
 
These positive definite coefficients will necessarily contain the dissipative transport coefficients but they have to satisfy only some inequality to maintain the positivity of the divergence.

  From this condition it seems that to maintain `the entropy condition'  no further `equality type' constraints are required other than the ones imposed by the `equilibrium condition'.
 
Below in section (\ref{sec:explicit}) we shall try to use our intuition to construct these higher order (in terms of $\omega$ expansion) terms in $J^\mu$.

\section{Construction of the higher order pieces in $J^\mu$}\label{sec:explicit}
Here we shall see how we can ensure the entropy positivity by adding new terms to the entropy current, we have already determined using the partition function. We would like to argue that such construction is always possible and the coefficients of these new terms will not be fixed by any transport coefficients though their ranges may be restricted by the dissipative terms in the stress tensor and the current.

Let us first decompose $\pi^{\mu\nu}$ and $j^\mu$ into two parts, `dissipative' and `non-dissipative'.
\begin{equation}\label{explicit1}
\begin{split}
\pi^{\mu\nu} = \pi_{(non-diss)}^{\mu\nu} + \pi_{(diss)}^{\mu\nu},
~~j^\mu = j_{(non-diss)}^\mu + j_{(diss)}^\mu\\
\end{split}
\end{equation}
where $\pi_{(non-diss)}^{\mu\nu}$ and $j_{(non-diss)}^\mu$ do not vanish in equilibrium (and hence will be called `non-dissipative'), $\pi_{(diss)}^{\mu\nu}$ and $j_{(diss)}^\mu$  vanish in equilibrium (and hence will be called `dissipative')\footnote{Such a decomposition is not unique, since we can always add some dissipative term to the non-dissipative ones without changing its nature}. Once the `equilibrium condition' is satisfied the divergence of the entropy current will contain only $\pi_{(diss)}^{\mu\nu}$ and $j_{(diss)}^\mu$. From equation \eqref{ent:divcan} it follows that in the expression of the divergence they will appear in the following way (through the divergence of the canonical part).
\begin{equation}\label{explicit2}
\begin{split}
\text{Divergence}\rightarrow &(j_{(diss)}^\mu u_\mu)(u.\partial\nu) -\left(\frac{u_\mu u_\nu \pi_{(diss)}^{\mu\nu}}{T^2} \right)(u.\partial T) -\left(\frac{P_{\mu\nu}\pi_{(diss)}^{\mu\nu}}{3T} \right)\Theta\\
 &+v_\mu j_{(diss)}^\mu + \left(\frac{u_\nu\pi_{(diss)}^{\mu\nu}}{T}\right)h_\mu -\left(\frac{\pi_{(diss)}^{\mu\nu}}{T}\right)\sigma_{\mu\nu} \\
 &+\text{dissipative terms arising from $D_\mu S^\mu$}
 \end{split}
\end{equation}

Next we observe that at first order in derivative expansion the exhaustive list of dissipative terms contains only the following elements.
$$(u.\partial T),~(u.\partial\nu),~\Theta, ~h_\mu,~v_\mu,~\sigma_{\mu\nu}$$
We shall collectively denote these terms as ${\mathfrak D}$.

Let us denote all higher order (in derivative expansion) dissipative terms by ${\mathfrak H}$. For convenience we shall further classify the elements of $\mathfrak{H}$ into three types.
\begin{enumerate}
\item ${\mathfrak H}^{(many)}$: Terms that have more than one factor  from the elements of ${\mathfrak D}$
\item ${\mathfrak H}^{(one)}$: Terms that have exactly one factor from the elements of ${\mathfrak D}$
\item ${\mathfrak H}^{(zero)}$: Terms that have no factor from the elements of ${\mathfrak D}$.
\end{enumerate}
 We shall use subscript to distinguish between different higher order dissipative terms. Naively we can construct any element of ${\mathfrak H}$ either by applying $D_\mu$ operator repeatedly on elements of ${\mathfrak D}$ or by applying $u^\mu D_\mu$ operator on any lower order non-dissipative scalar. We can also construct the composite dissipative terms (products of lower order terms) where at least one factor is an element of ${\mathfrak H}$ at lower order.

$\pi_{(diss)}^{\mu\nu}$ and $j_{(diss)}^\mu$ will contain terms of the form ${\mathfrak D}$ at first order in derivative expansion and of the form ${\mathfrak H}$ at higher order.

The full divergence of the entropy current will have the following type of scalars.
\begin{enumerate}
\item A scalar quadratic form in the elements of ${\mathfrak D}$.  We shall denote these terms together by ${\mathfrak Q}_1$
\item Scalars of type ${\mathfrak H}^{(many)}$.
\item Scalars of type ${\mathfrak H}^{(zero)}$.
\item Scalars of type ${\mathfrak H}^{(one)}$.
\end{enumerate}
If the `equilibrium condition' is satisfied, then by construction, our entropy current would be such that if we evaluate it on $\{\hat u^\mu,\hat T,\hat\nu\}$ it would be of order ${\cal O}(\omega^2)$. This should be true of all the four types of scalars that we have mentioned above. 

Now we shall analyse each of these four types of terms.
We shall start from $\mathfrak Q_1$. ${\mathfrak Q}_1$ is a quadratic form in the elements of ${\mathfrak D}$. Therefore it is second order in derivative expansion and it contains all the first order dissipative transport coefficients.
We can always diagonalise this quadratic form in the space of ${\mathfrak D}$ and write it as a sum of squares. 
Let us denote the eigen vectors as ${\mathfrak e}_i,~~i = 1,\cdots,6$.
So this part of the divergence schematically takes the following form.
$${\mathfrak Q}_1 = \sum_{i=1}^6\alpha_i {\mathfrak e}_i^2$$
In a generic case all eigenvalues $\alpha_i$s will be non-zero. And if we are interested in an
entropy current with non-negative divergence upto second order in derivative expansion we should constrain all these eigenvalues to be positive.

Next we come to $\mathfrak H^{(many)}$. The terms of the form ${\mathfrak H}^{(many)}$  are non zero only if ${\mathfrak Q}_1$ is non-zero and also  ${\mathfrak H}^{(many)}$ are always much smaller in magnitude than ${\mathfrak Q}_1$ in the regime of validity for the derivative expansion.
From here it immediately follows that the presence of ${\mathfrak H}^{(many)}$ type terms in the divergence can never determine the sign of the divergence. So if we are interested only about the constraints imposed by the `entropy condition' we can ignore these ${\mathfrak H}^{(many)}$ terms.

Finally  we shall analyse ${\mathfrak H}^{(zero)}$. We should note that
though ${\mathfrak H}^{(zero)}$ does not have any factors from the elements of $\mathfrak D$, it vanishes in equilibrium. We shall use this property to rewrite all ${\mathfrak H}^{(zero)}$ terms as a sum of ${\mathfrak H}^{(one)}$ and ${\mathfrak H}^{(many)}$ upto total derivative. In terms of equation we mean the following.
\begin{equation}\label{vvint}
{\mathfrak H}^{(zero)} = {\mathfrak H}^{(one)} +{\mathfrak H}^{(many)} + D_\mu {\mathfrak J}^\mu
\end{equation}
Where ${\mathfrak J}^\mu$ is a lower order current that also necessarily vanish in equilibrium. 
\newline
The intuition behind equation\eqref{vvint} is the following.
\newline
In our analysis the defining characteristic of `equilibrium' is that it is time independent. In other words $\partial_t$ or $\partial_0$ is a Killing vector for the equilibrium background. In the language fluid variables the Killing vector could be written as $\hat\xi^\mu =\frac{\hat u^\mu}{\hat T}$. According to the Killing equation,  in perfect equilibrium the symmetrized derivative of $\hat{\xi}^\mu$ vanishes, so does the Lie derivative of the gauge field in the direction of $\hat \xi^\mu$ (see \cite{loga1} for more detail).
\newline
 Now consider the symmetric tensor 
 $$S_{\mu\nu}=\left[D_\nu\left(\frac{u_\mu}{T}\right) +D_\mu\left(\frac{u_\nu}{T}\right)\right]$$
 and the vector
 $$V^\mu =\left[\left( \frac{u^\nu}{T} \right)D_\nu{\cal A^\mu} - {\cal A}^\nu D_\nu\left(\frac{u^\mu}{T}\right)\right]+{\cal A}^\nu S^\mu_\nu$$ 
   Clearly both $S_{\mu\nu}$  and $V^\mu$ vanish in equilibrium as a consequence of the Killing symmetry of the background\footnote{ The first term in $V^\mu$ is actually the expression for Lie derivative of the background gauge field in the direction of $\xi^\mu = \frac{u^\mu}{T}$. The second term is added to make $V^\mu$ gauge invariant. Since this second term is proportional to $S^{\mu\nu}$, it also vanishes in a time independent situation.}
 and both of them could be expressed in terms of the elements of $\mathfrak D$.
 \begin{equation}\label{elementd}
 \begin{split}
 S^{\mu\nu} &=2 u^\mu u^\nu\left(\frac{ u.\partial T}{T^2} \right)- \left(\frac{u^\nu h^\mu + u^\mu h^\nu}{T}\right) + \frac{2}{T}\left(\sigma^{\mu\nu} +\frac{\Theta}{3} P^{\mu\nu}\right)\\
 V^\mu &=-v^\mu - u^\mu (u.\partial\nu) 
 \end{split}
 \end{equation}
It is natural to expect that any non-trivial function of fluid variables and their derivatives that vanishes only in equilibrium must have some component of $S_{\mu\nu}$, $V^\mu$ or their derivatives as one of the factors.
  Then from equation \eqref{elementd} it follows that all $\mathfrak H^{(zero)} $ terms must contain factors where one or more derivatives are acting on the elements of $\mathfrak D$ with appropriate contractions so as to make it a scalar\footnote{This entire argument involving Killing symmetry was introduced to author by R. Loganayagam. Author sincerely thanks him for explaining it in detail to her.}.
  \newline
 Now we could always rearrange such a dissipative term and decompose it in the form of equation\eqref{vvint}. The idea is to shift the derivatives acting on the elements of $\mathfrak D$ to the other factors by repeatedly adding total derivative terms.
 To give a very simple example, consider a $\mathfrak H^{(zero)}$ scalar $X = \mathfrak a^\mu u^\nu D_\mu D_\nu \Theta$. We would like to rewrite it in form of equation \eqref{vvint}. The steps are as follows.
 \begin{equation}\label{exstep}
 \begin{split}
 X &= \mathfrak a^\mu u^\nu D_\mu D_\nu \Theta\\
 &=D_\mu\left(\mathfrak a^\mu u^\nu D_\nu \Theta\right) - D_\mu\left(\mathfrak a^\mu u^\nu\right) D_\nu \Theta\\
 &=D_\mu\left(\mathfrak a^\mu u^\nu D_\nu \Theta\right) -D_\nu\left[ D_\mu\left(\mathfrak a^\mu u^\nu\right) \Theta\right] + \Theta D_\nu\left[ D_\mu\left(\mathfrak a^\mu u^\nu\right)\right] 
 \end{split}
 \end{equation}
 In \eqref{exstep} the last line has the desired form of equation\eqref{vvint}. 

Equation \eqref{vvint} is the key equation that determines the inequalities to be satisfied to ensure the `entropy condition'.
In appendix (\ref{app:intarg}) we shall give some more arguments for why equation \eqref{vvint} seems to be true at all orders. The main point that needs further explanation is that we could often find dissipative term that apparently does not have any elements from ${\mathfrak D}$. For example, consider the scalar $(u.D)^2(D_\mu\mathfrak a^\mu)$, which manifestly vanish in equilibrium. Equation\eqref{vvint} requires a rewriting of such terms in a form where the derivatives are acting on the elements of 
$\mathfrak D$. In appendix (\ref{app:intarg}) we essentially tried to construct an explicit algorithm that we can use for such rewriting.
\\

Once we have rewritten  ${\mathfrak H}^{(zero)}$ as in equation \eqref{vvint} the full divergence will have terms of the form ${\mathfrak Q}_1$, ${\mathfrak H}^{(many)}$ and ${\mathfrak H}^{(one)}$ plus a total derivative piece (i.e. a term of the form $D_\mu {\mathfrak J}^\mu$).
We again  rewrite this ${\mathfrak H}^{(one)}$ as eigenvectors of ${\mathfrak Q}_1$ times ${\mathfrak H}^{(zero)}$ type higher order dissipative terms.
So this part of the divergence schematically takes the following form.
$${\mathfrak Q}_2 = \sum_{i=1}^6 \beta_i{\mathfrak e}_i {\mathfrak H}^{(zero)}_i$$
where ${\mathfrak H}^{(zero)}_i$ denotes different higher order dissipative terms.

At this stage
$$D_\mu J^\mu = \sum_{i=1}^6\alpha_i {\mathfrak e}_i^2+\sum_{i=1}^6 \beta_i{\mathfrak e}_i {\mathfrak H}^{(zero)}_i + D_\mu {\mathfrak J}^\mu$$
Now we shall redefine our entropy current by absorbing ${\mathfrak J}^\mu$ in $J^\mu$ . Then the divergence will take the following form.
\begin{equation}\label{bbba}
J^\mu\rightarrow (J^\mu - {\mathfrak J}^\mu)\Rightarrow D_\mu J^\mu = \sum_{i=1}^6\alpha_i {\mathfrak e}_i^2+\sum_{i=1}^6 \beta_i{\mathfrak e}_i {\mathfrak H}^{(zero)}_i
\end{equation}
If we want this divergence to be positive definite we have to add new terms to the entropy current so that its divergence produces terms of the form $\left[{\mathfrak H}^{(zero)}_i\right]^2$. 

This we can do order by order in derivative expansion.

Suppose we are interested at some fixed $n (>1)$th order in derivative expansion. Hence $\pi_{(diss)}^{\mu\nu}$ and $j_{(diss)}^\mu$ will contain $\mathfrak H$ of order  $n$. The same will happen for the divergence as given in RHS of equation \eqref{bbba}.

Now consider the scalars of the form $\left[{\mathfrak H}^{(zero)}_i\right]^2$. These are again  scalars of type  ${\mathfrak H}^{(zero)}$ at order $2n$ in derivative expansion. So using equation \eqref{vvint} we can again rewrite them as 

\begin{equation}\label{baba}
{\mathfrak H}^2\sim \sum_j \psi_j{\mathfrak e}_j {\mathfrak H}'_j + D_\mu {\mathfrak T}^\mu
\end{equation}
for some $(2n-1)$th order vector ${\mathfrak T}^\mu$ and $\psi_j$ are some arbitrary functions of scalar fluid variables like temperature and charges.  From here onwards we shall remove the superscript `$(zero)$' for convenience. All ${\mathfrak H}$'s that we have mentioned in the next few paragraphs are of ${\mathfrak H}^{(zero)}$ type.

In equation \eqref{baba} the ${\mathfrak H}$ s appearing in  LHS  are all different from the the ${\mathfrak H}'$ s appearing in the RHS.  We have an extra prime in RHS to denote this difference. In derivative expansion ${\mathfrak H}'$ s are of order $(2n-1)$. 

Therefore it follows that adding ${\mathfrak T}^\mu$ to the previous entropy current,  we can generate the terms of the form ${\mathfrak H}^2$ in the divergence. At this stage the divergence will be another quadratic form in ${\mathfrak e}$ and ${\mathfrak H}$ and a term of the form ${\mathfrak e}$ times ${\mathfrak H}'$ . 

To see whether these linear terms in ${\mathfrak H}'$ s could be further absorbed in a positive quadratic form we have to add terms to the entropy current so that its divergence produces $({\mathfrak H}')^2$ which is of order ${\cal O}(4n-2)$. This we can do exactly in the same way we did in the previous order. So we can continue this process infinite times.

But if we are interested in constraints at order ${\cal O}(n)$ we can simply ignore all these ${\mathfrak H}'$ s. We shall diagonalize the quadratic form between ${\mathfrak e}$ and ${\mathfrak H}$ and write the divergence as sum of squares.  Finally if we demand that  the coefficient of each of these square terms is non-negative, it will satisfy the `entropy condition'. Clearly the final constraints will only be in the form of inequalities. 

This shows that the entropy current that we have generated from the partition function can  always be extended to an entropy current with non negative divergence without any further `equality type' constraints on the transport coefficients.

Now a couple of comments about our argument.
\begin{enumerate}
\item We can arrive at this positive definite form of the divergence even without using any equations of motion or without fixing any frame. We expect that equations of motion will make the analysis simpler as it will reduce the number of independent terms.
\item We have started with an entropy current as determined from the partition function. At each order we added new terms to it so that the divergence could be written as a quadratic form in the space of independent dissipative terms upto the given order we are interested in.
\item If we are interested about the transport coefficients of order $n$ we have to add terms  of order $(2n -1)$ to the entropy current. The inequalities will definitely involve the coefficients of these new terms in the entropy current. So we should not consider them  as any physical constraint.
\item For $n=1$ we do not need to add new terms (in fact from explicit calculation it follows that we cannot add new term to the entropy current maintaining the positivity of the divergence \cite{secondorder},\cite{superfluident}). Therefore in derivative expansion this is the only order where the resultant inequalities will involve  the transport coefficients alone and not any coefficient from the entropy current.

\item From here it follows that the first order transport coefficients are the only ones that have to satisfy some inequalities to maintain the `entropy condition'.

\item Equation \eqref{vvint} is the key equation behind all these arguments. In appendix(\ref{app:intarg}) we give some more arguments about why we believe this equation to be true.

\end{enumerate}

\section{Ambiguities}\label{sec:ambiguity}
In the previous sections we have given a prescription of how, from partition function, we can construct one example of entropy current whose divergence is always non-negative. But we should emphasize that this is just one example and there is no claim of uniqueness for this construction. In this section we shall see what are the possible sources of this non-uniqueness.
\subsection{Terms with zero divergence}
Consider a covariant vector whose divergence vanishes identically. One could argue that such a vector must be of the form $R^\mu=D_\nu{\mathcal K}^{\mu\nu}$ where 
${\mathcal K}^{\mu\nu}$ is some antisymmetric tensor.  $[D_\mu R^\mu =0]$ simply because of the antisymmetry of ${\mathcal K}^{\mu\nu}$. At any given order in the entropy current we can always add terms of this form without affecting the `entropy condition'. In the language of differential forms this ambiguity is just equivalent to the freedom of adding an `exact' form  without affecting the exterior derivative. This is the same ambiguity that any conserved Noether current might have. We have already noted in section(\ref{sub:step3}) that our construction of entropy current has similarities with the construction of Noether current and Wald entropy in the theories of higher derivative gravity (\cite{Wald},\cite{Iyer}). The ambiguities that we are describing here are also similar to the ones found in the context of gravity. See \cite{Iyer} for more detailed explanation on this. 

 Now we would like to see what these ambiguities translate to when we are using the `equilibrium partition function' to determine the entropy current.
We should note that the time component of these vectors ($R^0$) could well be non-zero in equilibrium and therefore naively they contribute to the total entropy. 
We have also seen that in equilibrium the `non-canonical' part of the total entropy is exactly equal to the derivative corrections to the partition function. Hence we should be able to relate $R^0$ to some terms in the partition function itself.
On the other hand, it is obvious that from the point of view of  `entropy condition', $R^\mu$ cannot play any role to determine the constraints on the transport coefficients. This would be possible only if $R^0$ evaluates to a total derivative in equilibrium. By explicit evaluation we can see that this is true.
\begin{equation}\label{totalderi}
\begin{split}
\sqrt{G}R^0|_{equilibrium} &=\partial_i\left[\sqrt{G}\mathcal K^{0i}\right]
=\sqrt{g}\nabla_i\left[e^{\sigma}\mathcal K^{0i}\right]
\end{split}
\end{equation}
Equation \eqref{totalderi} shows that any term in the entropy current with identically vanishing divergence could be related to a total derivative term in the partition function. 
\newline
Next we would like to see whether the converse of the above statement is true, that is, whether for every total derivative piece in the partition function we could write a term of the form of $R^\mu$ in the entropy current.

Consider a total derivative piece in the partition function 
$$W_{total~derivative} = \int \sqrt{g} \nabla_i \hat X^i$$
According to the prescription described in section(\ref{sec:general}) $W_{total~derivative}$ will generate the following entropy current $\hat S_{total~derivative}^\mu$ near equilibrium.
\begin{equation}\label{enttoal}
\begin{split}
\hat S_{total~derivative}^0 &= e^{-\sigma} \nabla_j \hat X^j,~~
\hat S_{total~derivative}^i =  e^{-\sigma}\partial_0\left(\sqrt{g} X^i\right)
\end{split}
\end{equation}
Now we have to find a vector $S_{total~derivative}^\mu$ such that on $\{\hat u^\mu,\hat T,\hat\nu\}$ it reduces to $\hat S_{total~derivative}^\mu$.
The first step would be to covariantize $\hat X^i$. Suppose $X^\mu$ is the vector that evaluates to $\hat X^i$ in equilibrium\footnote{It is possible that $\hat X^i$ is not invariant under KK gauge transformation (coordinate transformation that changes $t\rightarrow t'= t + f(\vec x)$). Suppose $\hat X^i$ is of the form 
$$\hat X^i\sim a_j \hat C^{ji}$$ 
 Here if $\hat C^{ji}$ is an antisymmetric but KK invariant tensor, then $(\nabla_i X^i)$ would be an allowed term in the partition function. This is because under KK gauge transformation it generates only a boundary term. See \cite{equipart},\cite{equitarun},\cite{equiloga},\cite{equikristan},\cite{loga1},\cite{loga2} for examples.
 
 In such a case it is not possible to find a covariant vector $X^\mu$ whose $i$ component is $X^i$. Instead we have to covariantize $\hat C_{ij}$ into an antisymmetric tensor $C_{\mu\nu}$ with both indices projected in the direction perpendicular to $u^\mu$. Then $D_\nu C^{\mu\nu}$ would be the desired term in the entropy current. Unlike the other terms in the entropy current, this particular one will have non-zero space component even when evaluated on $\{\hat u^\mu,\hat T,\hat\nu\}$ in perfect equilibrium.}. Then we should construct $S_{total~derivative}^\mu$ in the following way.
\begin{equation}\label{covsmu}
\begin{split}
S_{total~derivative}^\mu =D_\nu (u^\mu X^\nu - u^\nu X^\mu)
\end{split}
\end{equation}
To check we can evaluate $S_{total~derivative}^\mu$ explicitly on $\{\hat u^\mu, \hat T,\hat\nu\}$.

\begin{equation}\label{explev0}
\begin{split}
\hat S_{total~derivative}^0
=~&D_\nu(\hat u^0\hat X^\nu - \hat u^\nu \hat X^0)
=~\frac{1}{\sqrt G}\partial_i\left[\sqrt{G}e^{-\sigma}\hat X^i\right]\\
=~&e^{-\sigma}\nabla_i\hat X^i
\end{split}
\end{equation}
\begin{equation}\label{explevi}
\begin{split}
\hat S_{total~derivative}^i
=~&D_\nu(\hat u^i\hat X^\nu - \hat u^\nu \hat X^i)
=~-\frac{1}{\sqrt G}\partial_0\left[\sqrt{G}e^{-\sigma}\hat X^i\right]\\
=~&-\frac{e^{-\sigma}}{\sqrt g}\partial_0\left[\sqrt{g} \hat X^i\right]\\
\end{split}
\end{equation}
Thus we see that for every total derivative term in the partition function we could construct an entropy current whose divergence vanish identically. In other words there is a one to one correspondence between every total derivative piece in the partition function and the zero divergence term in the entropy current, both having no impact on the transport coefficients.

\subsection{Ambiguity in covariantizing $\hat J^\mu$}
As we have explained before, from the partition function we could determine a current $\hat J^\mu$.
But this current is highly non-covariant since here the variation in time is much slower than the variation in space and so time and space are treated on a different footing. We have defined the covariant current $J^\mu$ to be such that it reduces to $\hat J^\mu$ upto order ${\cal O}(\omega^2)$ when evaluated on $\{\hat u^\mu,\hat T,\hat \nu\}$. Clearly this construction is not unique since we can always add any term to $J^\mu$ that is of order ${\cal O}(\omega^2)$ on $\{\hat u^\mu,\hat T,\hat \nu\}$, without affecting the above requirement. This is the part of the entropy current that could not be determined using the partition function. However in section(\ref{sec:explicit}) we have seen how any such order ${\cal O}(\omega^2)$ term in the entropy current could be handled by adding new higher order terms and hence the presence of such terms do not give any new constraints on the transport coefficients.

One way to characterize this ambiguity would be to write out the most general possible form of the entropy current upto some given order. This will simply be determined by the symmetry of the system. Then we should evaluate it on $\{\hat u^\mu,\hat T,\hat \nu\}$ and compare the answer with $\hat J^\mu$ we have already derived from the partition function. There will be some part in the most general form of the entropy current that could not be fixed by this method. This is a genuine ambiguity which might even have a physical significance. It probably says that the entropy current is a concept that is uniquely defined only in solutions that are very close to being static. If one is far away from a stationary state, even to define it properly we need something more than the physical principle of local entropy production that we are using here throughout.

 \section{One `toy' example}\label{sec:example}
 In this section we shall implement  the above algorithm to construct the entropy current for a very simple system. We should emphasize that this example is designed just to explicitly show how our construction works in a simple situation and it does not have any other physical significance. Please, see \cite{Toapp} for application of this method to a physically relevant and more complicated case of non-anomalous charged fluid at second order. 
 
  Consider  uncharged fluid at second order in derivative expansion.  The  partition function for such a case generically will have three independent terms multiplied by three arbitrary coefficients which are functions of temperature \cite{equipart},\cite{equipartamos}. But for simplicity here we shall set two of these three coefficients to zero. There is no physical reason for doing this and most likely it will result in a very unnatural set of transport coefficients. However, as explained above, this simple case will serve the technical purpose of showing how effectively the formalism developed so far, could be implemented.
  
So here the equilibrium partition function has a single term at second order in derivative expansion.   This is a special case of the fluid considered in \cite{secondorder},\cite{equipart},\cite{equipartamos}. Therefore the entropy current will also be a special case of what has been constructed in \cite{secondorder}.
 
 Suppose, the partition function is given as follows.
\begin{equation}\label{ex:part}
\begin{split}
W &= \int \sqrt{g}\left[\frac{p(\hat T)}{\hat T} + K(\hat T)(\nabla \hat T)^2\right]\\
L_{(0)}&=\frac{p(\hat T)}{\hat T},~~L_{(1)}=0,~~~L_{(2)} =K(\hat T)(\nabla \hat T)^2
\end{split}
\end{equation}
 The partition function does not depend on the gauge field. So clearly for this simple case the current is zero. Substituting equation \eqref{ex:part} in equation \eqref{stc} we can read off $\hat E$, $\hat P$ and also the various components of $\hat \pi^{\mu\nu}$.
 \begin{equation}\label{ex:equi}
 \begin{split}
 &\hat E = \hat T \left(\frac{d p}{d\hat T}\right) - p,~~\hat P = p\\
& \frac{\hat u^\mu \hat u^\nu \hat\pi_{\mu\nu} }{\hat T^2} = - \left(\frac{d K}{d\hat T}\right)(\nabla \hat T)^2-2K \nabla^2\hat T\\
&\frac{\hat P^i_\mu u_\nu \hat\pi^{\mu\nu}}{\hat T^2} = 0\\
&\frac{\hat P_{i\alpha} \hat P_{j\beta} \hat\pi^{\alpha\beta} }{\hat T}  = -  2K\left(\nabla_i \hat T\nabla_j \hat T - \frac{g_{ij}}{2}(\nabla \hat T)^2\right)
 \end{split}
 \end{equation}
 The `equilibrium condition will be satisfied provided 
 \begin{equation}\label{ex:eqst}
 \begin{split}
\lim_{\omega\rightarrow0}\pi^{\mu\nu} =\hat\pi^{\mu\nu}
 \end{split}
 \end{equation}
 where different components of $\hat\pi^{\mu\nu}$ are determined in terms of the partition function through equation \eqref{ex:equi}.
 
 Using the rules of thermodynamics we can derive the total entropy from the partition function.
\begin{equation}\label{ex:thermo}
\begin{split}
\text{Total entropy}& = S_T\\
&= W + T_0 \left(\frac{\partial W}{\partial T_0}\right)\\
&=\int \sqrt{g}\left[ \frac{d p}{d\hat T} +K(\nabla \hat T)^2  + 2K( \nabla\hat T)^2 + \hat T \left(\frac{d K}{d\hat T}\right)(\nabla \hat T)^2\right]\\
&=\int \sqrt{g}\left[ s + L_{(2)} -2\hat TK \nabla^2\hat T +2 \nabla_i(K\hat T  \nabla^i T)- \hat T \left(\frac{d K}{d\hat T}\right)(\nabla \hat T)^2\right]\\
&=\int \sqrt{g}\left[ s + L_{(2)} + \frac{\hat u^\mu \hat u^\nu \hat\pi_{\mu\nu} }{\hat T} + 2\nabla_i(K\hat T \nabla^i T)\right]
\end{split}
\end{equation}
In equation \eqref{ex:thermo} we have used equations \eqref{ex:part} and \eqref{ex:equi} and also we have used $s = \text{entropy density} =\frac{dp}{d\hat T}$

Now the total entropy is related to the integration of the zero component of the entropy current over some space-like slice.
\begin{equation}\label{ex:j1}
S_T = \int \sqrt{G} \hat J^0 = \int \sqrt{g} e^{\sigma} \hat J^0 
\end{equation}
Comparing equation \eqref{ex:j1} with \eqref{ex:thermo} we can determine $\hat J^0$ upto total derivatives. We find
\begin{equation}\label{ex:j2}
\hat J^0 = e^{-\sigma}\left[ s  - \frac{\hat\pi^0_0 }{\hat T}+ L_{(2)} + \nabla_i\mathfrak K^i\right]
\end{equation}
where $\mathfrak K^i = 2(K\hat T \nabla^i T)$ is some vector constructed out of background and we have used that  $\hat \pi^i_0$ is zero for our case.

Now we have to compute the time derivative of $\hat J^0$.

We shall first process the last term in equation \eqref{ex:j2} or \eqref{ex:thermo}. This term, being a total derivative is a purely boundary term to begin with  and therefore will remain so after $\partial_0$ acts on it. In other words the time derivative of the total derivative terms in the entropy can be trivially rewritten as a boundary term and this will not require any constraint to be satisfied in the bulk. Therefore in what follows we shall simply ignore the total derivative pieces in the expression of total entropy in equilibrium or in the expression of $\hat J^0$.  

So for our purpose we shall write $\hat J^0$ and $S_T$ as
\begin{equation}\label{ex:j3}
\begin{split}
\hat J^0 &= e^{-\sigma}\left[ s  - \frac{\hat\pi^0_0 }{\hat T}+ L_{(2)}\right]\\
S_T&
=\int \sqrt{g}\left[ s  - \frac{\hat\pi^0_0 }{\hat T}+ L_{(2)}\right]\\
\end{split}
\end{equation}
We can see that equation \eqref{ex:j3} is a special case of equation \eqref{desfo} and therefore we can directly apply equation \eqref{candiv}, \eqref{lasttwo}, \eqref{finalcandiv} and finally equation \eqref{explicit} for the time derivative of the first two terms of equation \eqref{ex:j3}. We get the following.
\begin{equation}\label{time2}
\begin{split}
\partial_0\left(\sqrt{g}s-\sqrt{g}\frac{ \hat\pi^0_0}{\hat T}\right) &=\sqrt{g}
\left(\frac{\hat\pi_{ij}\partial_0 g^{ij}}{2\hat T} -e^{-2\sigma}\frac{\hat\pi_{00}\partial_0\hat T}{2\hat T^2}\right) +\partial_i\left(\sqrt{g}\frac{ \tilde\pi^i_0}{\hat T}\right) + {\cal O}(\omega^2)
\end{split}
\end{equation}

Now we are going to compute the time derivative of the third term in equation \eqref{ex:j3}.
\begin{equation}\label{ex:third}
\begin{split}
&\frac{1}{\sqrt g}\partial_0\left(\sqrt{g}L_{(2)}\right) = \frac{1}{\sqrt g}\partial_0\left[\sqrt{g}K(\nabla\hat T)^2\right]\\
&=\left[-K\frac{(\nabla\hat T)^2}{2}g^{ij}+ K(\nabla_i \hat T)(\nabla_i \hat T)\right](\partial_0g^{ij}) +  (\nabla \hat T)^2\partial_0 K
+\left[ 2K(\nabla^i \hat T)(\nabla_j\partial_0 \hat T)\right]\\
&=K\left[-\frac{(\nabla\hat T)^2}{2}g^{ij} + (\nabla_i \hat T)(\nabla_j \hat T)\right](\partial_0g^{ij})
 - 2K(\nabla^2 \hat T)(\partial_0 \hat T)-\left(\frac{dK}{d\hat T }\right)(\nabla \hat T)^2 (\partial_0 \hat T)\\
&~~~~~~+\frac{1}{\sqrt g}\partial_i \bigg[2K\sqrt{g}g^{ij}(\partial_j \hat T)(\partial_0\hat T)\bigg] \\
&=-\sqrt{g}
\left(\frac{\hat\pi_{ij}\partial_0 g^{ij}}{2\hat T} -e^{-2\sigma}\frac{\hat\pi_{00}\partial_0\hat T}{2\hat T^2}\right) +\partial_i \bigg[2K\sqrt{g} g^{ij}(\partial_j \hat T)(\partial_0\hat T)\bigg]
\end{split}
\end{equation}
In the last line we have used equation \eqref{ex:equi} which is a consequence of the `equilibrium condition'. Now we see that if we combine equations \eqref{time2} and \eqref{ex:third}, the part involving the equilibrium stress tensor cancels and we get pure boundary terms.

From equations\eqref{time2} and \eqref{ex:third} it follows that 
\begin{equation}\label{ex:bterm}
\begin{split}
\partial_0 S_T =
\int \sqrt{G}\bar\nabla_j\left[ 2K(\bar\nabla^j \hat T)(\hat u^\mu\partial_\mu\hat T) + \frac{ \hat u^\nu\tilde\pi^j_\nu}{\hat T}\right] + {\cal O}(\omega^2)\\
\end{split}
\end{equation}
From equation \eqref{ex:bterm} and \eqref{ex:j3} we can read off the time and space component of $\hat J^\mu$
\begin{equation}\label{ex:j4}
\begin{split}
\hat J^0 &=e^{-\sigma}\left[ s  - \frac{\hat\pi^0_0 }{\hat T}+ L_{(2)}\right]\\
\hat J^i &=- 2K(\bar\nabla^i \hat T)(\hat u^\mu\partial_\mu\hat T) - \frac{ \hat u^\nu\tilde\pi^i_\nu}{\hat T}\\
\end{split}
\end{equation}

Now we have to construct a current $J^\mu$ that will reduce to $\hat J^\mu$ when evaluated on $\{\hat u^\mu,\hat T,\hat\nu\}$ upto order ${\cal O}(\omega)$. One obvious choice in our case would be the following\footnote{To
 covariantize $\hat J^\mu$ we have simply replaced all $\nabla_i$ by $D_\mu$. Ideally we should replace $\nabla_i$ by $ P_\mu ^\alpha D_\alpha$. But the difference is of order ${\cal O}(\omega^2)$ in this case and therefore it does not matter. This step also indicates the non-uniqueness of the covariant form of the entropy current}.
$$J^\mu = \left[s  u^\mu-   \frac{ u^\nu\pi^\mu_\nu}{ T}\right]+\left[K {\cal G}^{\alpha\beta}(D_\alpha  T)(D_\beta  T)  u^\mu-2K{\cal G}^{\mu\beta} (D_{\beta}  T)( u^\alpha D_\alpha T)\right] +\tilde S^\mu$$
where $\tilde S^\mu|_{\{\hat u^\mu,~\hat T,~\hat\nu\}} = {\cal O}(\omega^2)$ .
We should note that this is not any unique choice since from partition function nothing is determined beyond order ${\cal O}(\omega)$.

Now the choice of $\tilde S^\mu$ will depend on the dissipative part of the stress tensor and the current. But for any form of the dissipative part we should be able to choose at least one $\tilde S^\mu$ such that the divergence is positive definite upto the required order in derivative expansion provided dissipative transport coefficients satisfy some inequalities.

Now we shall write the most general form of the stress tensor consistent with the partition function.
 If we just use symmetry at second order there could be 15 independent transport coefficients \cite{secondorder}. 7 of them are dissipative i.e. they vanish in equilibrium. Rest 8 are constrained by the partition function. \cite{equipart} and \cite{equipartamos} have done the analysis for the most general partition function for uncharged fluid. But here we have chosen a very special and simple form for the partition function. So the non-dissipative part of the stress tensor will be a special case of what is presented in \cite{equipart} or \cite{equipartamos}. However for the dissipative part we shall choose the most general form. At second order it will have 7 transport coefficients.
\begin{equation}\label{ex:expl}
\begin{split}
\pi_{\mu\nu} = &~~\alpha (u.\partial T)u_\mu u_\nu  +\eta \sigma_{\mu\nu} + \zeta\Theta P_{\mu\nu}\\
& +\tau(u.D)\sigma_{\langle\mu\nu\rangle} + \lambda_1 \sigma^\alpha_{\langle\mu}\omega_{\nu\rangle\alpha}+ \lambda_2 \sigma^\alpha_{\langle\mu}\sigma_{\nu\rangle\alpha} +\lambda_3\Theta \sigma_{\mu\nu}
+\left[\zeta_1 (u.D)\Theta + \zeta_2 \Theta^2 + \zeta_3\sigma^2\right]P_{\mu\nu}\\
& -2KT (D_{\langle\mu}T)(D_{\nu\rangle}T) +\frac{P_{\mu\nu}}{3}\left[P^{\alpha\beta}KT(D_\alpha T )(D_\beta T)\right]\\
&- 2u_\mu u_\nu T^2K\left[\frac{1}{2K}\frac{dK}{dT} (DT)^2+D^2T - {(u.Du }^\alpha) D_\alpha T\right]\\
&+\text{Higher order terms}
\end{split}
\end{equation}
Here  for any tensor $A_{\langle\mu\nu\rangle}$ implies the following.
$$A_{\langle\mu\nu\rangle} = P_\mu^\alpha P_\nu^\beta\bigg[\left(\frac{A_{\alpha\beta} + A_{\beta\alpha}}{2} \right)- {\cal G}_{\alpha\beta}\left(\frac{P^{\theta\phi}A_{\theta\phi}}{3}\right)\bigg]$$
In equation \eqref{ex:expl} the all terms in the first two lines vanish in equilibrium\footnote{Equation \eqref{ex:prev} is actually a redundant description of the stress tensor since we can always absorb some of the transport coefficients in a frame redefinition. For example, in Landau frame the first terms in the first and the fourth line will be removed by frame redefinition. But here we are not going to fix any frame and we have allowed this redundancy. If we can construct an entropy current whose divergence is non negative in one frame it will remain so in all other frames since frame redefinition is just a rewriting of the same expression in a different language}. The rest evaluates to something non-zero in a time independent situation and we can explicitly check that \eqref{ex:expl} is consistent with the `equilibrium condition' i.e.. equation \eqref{ex:part}.

Using equation \eqref{ex:expl} we can explicitly compute the divergence of $J^\mu$.
\begin{equation}\label{ex:prev}
\begin{split}
&D_\mu J^\mu \\= &-\frac{\alpha}{T^2}(u.\partial T)^2-\frac{\zeta}{T}\Theta^2 -\frac{\eta}{T}\sigma^2 \\
& - K(u.\partial T)^2\Theta-\frac{\sigma^{\mu\nu}}{T}\left[\lambda_1 \sigma^\alpha_{\mu}\omega_{\nu\alpha}+ \lambda_2 \sigma^\alpha_{\langle\mu}\sigma_{\nu\rangle\alpha} +\lambda_3\Theta \sigma_{\mu\nu}\right] 
-\frac{\Theta}{T}\left[ \zeta_2 \Theta^2 + \zeta_3\sigma^2\right]\\
&-\frac{\tau}{T}\sigma^{\mu\nu}(u.D)\sigma_{\mu\nu}- \frac{\zeta_1}{T}\Theta (u.D)\Theta+ D_\mu \tilde S^\mu
\end{split}
\end{equation}
See appendix (\ref{derivation}) for a derivation of this equation.
\newline
In equation \eqref{ex:prev} the first three terms together form the ${\mathfrak Q}_1$ (quadratic form in the space of ${\mathfrak D}$ as we have defined in section \ref{sec:explicit}). For this very special case ${\mathfrak Q}_1$ is diagonal here to begin with.

All the scalars appearing in the second line are of ${\mathfrak H}^{(many)}$ type. They are always suppressed compared to the first line in equation \eqref{ex:prev} in the derivative expansion. Therefore these terms can never change the sign of the divergence as we have explained in section (\ref{sec:explicit}). The first two terms in the last line are of ${\mathfrak H}^{one}$ type and they can potentially violate the `entropy condition'.

Now if we can choose an $\tilde S^\mu$ such that its divergence generates terms of the form   $[(u.D)\sigma^{\mu\nu}][(u.D)\sigma_{\mu\nu}]$ and $[(u.D)\Theta]^2$, then we can easily express the RHS of equation \eqref{ex:prev} as sum of squares.

We observe the following
\begin{equation}\label{ex:statep}
\begin{split}
k_1[(u.D)\sigma^{ab}][(u.D)\sigma_{ab}] &=D_\mu\left[k_1 u^\mu\sigma_{ab}(u.D)\sigma^{ab}\right]-\sigma_{ab}D_\mu\left[k_1u^\mu (u.D)\sigma^{ab}\right]\\
k_2 [(u.D)\Theta]^2 &=D_\mu \left[k_2 u^\mu\Theta (u.D)\Theta\right]- \Theta D_\mu\left[ k_2 u^\mu (u.D)\Theta\right]
\end{split}
\end{equation}
Equation \eqref{ex:statep} is an example of the statement we made in equation \eqref{vvint}.
Clearly if we choose $\tilde S^\mu$ as
$$\tilde S^\mu = k_1 u^\mu\sigma_{ab}(u.D)\sigma^{ab} +k_2 u^\mu\Theta (u.D)\Theta$$
we shall generate the required terms.
With this choice of $\tilde S^\mu$ the full divergence of the entropy current takes the following form.
\begin{equation}\label{divagain}
\begin{split}
D_\mu J^\mu &=  - \frac{\alpha}{T^2}(u.\partial T)^2 - K(u.\partial T)^2\Theta \\
&- \frac{\zeta}{T}\left[\Theta
+ \frac{\zeta_1}{2\zeta} (u.D)\Theta\right]^2
+ \left[\frac{\zeta_1^2}{4 T\zeta} + k_2\right] \left[(u.D)\Theta\right]^2\\
&- \frac{\eta}{T}\left[\sigma_{\mu\nu}
+ \frac{\tau}{2\eta} (u.D)\sigma_{\mu\nu}\right]^2
+ \left[\frac{\tau^2}{4 T\eta} + k_1\right] \left[(u.D)\sigma_{\mu\nu}\right]^2\\
 &~+\sigma_{ab}D_\mu\left[k_1u^\mu (u.D)\sigma^{ab}\right] + \Theta D_\mu\left[ k_2 u^\mu (u.D)\Theta\right]\\
\end{split}
\end{equation}

In equation \eqref{divagain} the sum of the first 6 terms will always be positive provided the following inequalities are satisfied.
\begin{equation}\label{inequal}
\begin{split}
\frac{\zeta}{T}\leq0,~~\frac{\alpha}{T^2}\leq0,~~\frac{\eta}{T}\leq0,~~\left[\frac{\tau^2}{4 T\eta} + k\right]\geq0,~~\left[\frac{\zeta_1^2}{4 T\zeta} + k_2\right]\geq0
\end{split}
\end{equation}
The last two terms in \eqref{divagain} at this stage could have any sign, but we can further modify $\tilde S^\mu$ by adding 5th order terms in derivative expansion so that this term can again be absorbed in sum of squares. This process could go on indefinitely. However if we are interested only upto second order in derivative expansion we can truncate the process here and ignore all the higher order terms including the ones appearing in the last two lines of equation \eqref{divagain}.
\newline
From equation \eqref{inequal} we could also see that only the first order transport coefficients are the ones that have to satisfy genuine inequalities to ensure the `entropy condition' as claimed in section (\ref{sec:explicit}).

\section{Conclusion}\label{sec:conclude}
In this note we have reasoned why, in a general context,  the existence of equilibrium partition function and the existence of an entropy current lead to similar constraints for the transport coefficients. We showed the equivalence by explicit construction of one entropy current with non-negative divergence, starting from the partition function of the system.

So our starting assumption is that the  equilibrium partition function exists in a static background and the equilibrium values for the stress tensor and the current could be generated from this partition function. The fact that our general entropy current must be conserved in perfect equilibrium, and also it must integrate to the same total entropy we determined from partition function, gives the first set of constraints on its possible form.

Next we introduced a very slow time dependence in the background. We argued that in such cases the net entropy production must vanish in the bulk of the space, that is the entropy current must be adiabatically conserved. Using this fact we could further constrain the entropy current of the system in terms of the variation of the partition function on the boundary. In addition, we also see that an adiabatically conserved entropy current is possible only if the equilibrium values of the stress tensor and the charge current are consistent with the partition function.

Finally we have seen that we can always add some new terms (order by order in derivative expansion) to this entropy current so that its divergence is non-negative on any solution of fluid equations. The coefficients of these new terms are not fixed in terms of any transport coefficients and they are allowed to take a range values without violating the condition of local entropy production. Also only at first order in derivative expansion, some transport coefficients have to satisfy some inequalities to ensure the local entropy production. There are no other `inequlaity' type constraints for higher order transport coefficients.

So in summary our observation are the following.
\newline
 For a fluid system if an equilibrium partition function exists in a static background and if the first order dissipative transport coefficients satisfy some inequalities, then we can always construct an entropy current whose divergence is non negative on any consistent fluid flow. 
\newline
Part of this entropy current is completely determined in terms of the partition function. The coefficients for the rest of the terms are not fixed. We could choose them from a range of values. These ranges might depend on some dissipative transport coefficients.

The main assumptions that go into these arguments are as follows.
\begin{enumerate}
\item In a situation with slow time dependence in the background there exists at least one solution that approximately follows the previous equilibrium, now slowly shifting with time.

\item For such a solution any regular function of fluid variables (including the stress tensor and the current) has an analytic expression around the zero frequency limit for any arbitrary space variation.

\end{enumerate}

It is necessary to rigorously justify all these  assumptions so that our statement about the relation between the entropy current and the partition function is properly proved.

Also we have restricted our analysis strictly to non-anomalous fluid. However, we know that the presence of anomaly completely determines some of the transport coefficients. The entropy current is also already determined for these anomalous cases. It would be nice to generalize our algorithm to account for anomaly as well and rederive the known entropy currents using this method in all dimensions. 

It seems that if we do a linear fluctuation analysis around some equilibrium configuration of the fluid system, the signs of the leading dissipative terms will control the dynamical stability. The leading dissipative terms are nothing but the ones appearing in the stress tensor and current at first order in derivative expansion. And the signs that we predict from the inequalities generated from the `entropy condition' are same as required for the stability. These dissipative transport coefficients act like `damping terms' in the fluid equations. It is possible that the type of adiabatic solutions that we have assumed throughout this note exists for a range of initial conditions only if these inequalities are satisfied\footnote{Author sincerely thanks Shiraz Minwalla for explaining this possibility to her.}. Therefore in short, it seems that the existence of equilibrium and its stabilty is enough to ensure the existence of an entropy current with non-negative divergence. It would be nice to make this connection more rigorous by doing a complete linear stability analysis for the most general equilibrium fluid configuration on an arbitrary static background.

 We know that each of the fluid transport coefficients measure certain retarded correlators of the stress tensor and the currents in the long wavelength limit \cite{Schaefer:2014awa},\cite{Moore:2010bu}, \cite{Moore:2012tc}. As an effective description of some underlying quantum field theory we expect these correlators  to obey several symmetry conditions in the fluid limit. Equilibrium partition function consistently encodes all zero frequency correlators, but it says nothing about finite frequencies. For example, we know that time reversal symmetry of the correlators generate `Onsagar relations' among the dissipative transport coefficients\cite{landau}, which, neither the `equilibrium condition' nor `the entropy condition' could capture. It would be very nice if we could also derive these `Onsagar relations' from such near equilibrium analysis.
 
 Our final goal would be to write a single principle from which we can dertermine all the constraints that a physically consistent fluid must satisfy. In other words we want a principle that has the information about all correlators of the system in the long wavelength limit. These aspects are being studied for long time \cite{Kovtun:2012rj},\cite{Kovtun:2014hpa}.  We have seen that it is possible to write an `action' for fluid systems, at least in the non-dissipative cases \cite{Dubovsky:2011sj},\cite{Jensen:2013vta}. It would be interesting to see if all these different approaches could finally fit into a single line of thought.

Finally it would be very interesting to see what all these equivalences say about a gravity system that is dual to some fluid using the `fluid-gravity map'. We have already seen that our method has some similarities with the construction of Wald entropy in the higher derivative theories of gravity \cite{Wald}, \cite{Iyer}. We do not yet know whether this Wald entropy satisfies the second law of thermodynamics in the most general situation. It would be nice if our method could somehow be extended in the direction of gravity theories, where the $\alpha'$ correction of string theory could be treated on the same footing as the derivative expansion in fluid dynamics.
\acknowledgments
I would like to thank Shiraz Minwalla for providing guidance at every stage.
I thank Nabamita Banerjee and Suvankar Dutta for collaboration in the initial part of this work and for useful and stimulating discussions. I would like to thank Sachin Jain, Kristan Jensen, R. Loganayagam, Shiraz Minwalla, Mukund Rangamani and Amos Yarom for reading the draft and for very useful discussions and comments. Finally I would like to acknowledge our debt to the people of
India for their generous and steady support to research in the basic science.

\appendix

\section{Arguments for equation (6.3) }\label{app:intarg}
Here we shall give some intuitive arguments about 
why equation \eqref{vvint} is true. In particular  we shall try to chalk out a method that can be used to reduce any $\mathfrak H^{(zero)}$ term to the useful form, described in section(\ref{sec:explicit}). This is the form which could easily be re-arranged to give equation\eqref{vvint}.

Suppose  we have a dissipative scalar $S $ at order $n$ in derivative expansion. We would like to show that it could always be expressed in a particular way so that it will have at least one factor where the $D_\mu $ operator repeatedly acts (with appropriate contractions) on the elements of ${\mathfrak D}$.

To show this we shall first evaluate the scalar on $\{\hat u^\mu, \hat T, \hat\nu\}$ in the background as given in equation \eqref{equinot}.

In general any scalar will admit a power series expansion in terms of $\omega$.
$$S|_{\{\hat u^\mu, \hat T, \hat\nu\}} = \sum_{k=k_i}^{k={k_f} }\omega^k s_k, ~~~k_i>0$$
Now $\omega^{k_i} s_{k_i}$ (i.e. the leading term in the expansion) could have only the following terms.
\begin{enumerate}
\item Factors of $\partial_0 g_{ij}$, $\partial_0 a_i$, $\partial_0 A_i$, $\partial_0 \hat T$ or $\partial_0 \hat\nu$ operated by further
$\partial_i^k$ or $\partial_0^k$  operators.
\item The factors described above, could also be multiplied by terms without any $\partial_0$ operator
\end{enumerate}
In the expression of $\omega^{k_i} s_{k_i}$ we shall apply the reverse of equation \eqref{explicit} and shall  do the following replacement.
\begin{equation}\label{replacement1}
\begin{split}
&\text{Dissipative terms:}\\
&e^{-\sigma}\partial_0 g_{ij}\rightarrow -2\left(\sigma_{\mu\nu}+\frac{\Theta}{3} P^{\mu\nu}\right),~~~
e^{-\sigma}\partial_0 a_{i}\rightarrow -T h_\mu,
~~e^{-\sigma}\partial_0 A_{i}\rightarrow -T\left(v_\mu-\nu h_\mu\right)\\
&e^{-\sigma}\partial_0 \hat T\rightarrow (u.\partial T),~~
e^{-\sigma}\partial_0 \hat \nu\rightarrow (u.\partial \nu)\\
&\text{Non-dissipative terms:}\\
&f_{ij}=\left[\partial_i\hat a_j - \partial_j \hat a_i\right]\rightarrow -2TP^\alpha_\mu P^\beta_\nu(D_\alpha u_\beta - D_\beta u_\alpha)\\
&\left[\partial_i A_j - \partial_j A_i\right]\rightarrow P^\alpha_\mu P^\beta_\nu\left[{\cal F}_{\alpha\beta} +\frac{\nu}{2T}(D_\alpha u_\beta - D_\beta u_\alpha)\right]\\
&\partial_i \hat T \rightarrow P^\alpha_\mu\partial_\alpha T,~~\partial_i \hat \nu \rightarrow P^\alpha_\mu\partial_\alpha \nu\\
\\
&\text{Terms at zero derivative order:}\\
&e^{\sigma} \rightarrow \frac{T_0}{T},~~~~\frac{A_0}{T_0} \rightarrow \nu,~~~~g_{ij}\rightarrow P_{\mu\nu}
\end{split}
\end{equation}
and also individual derivatives as 
$$\partial_0\rightarrow u^\mu D_\mu, ~~~\partial_i\rightarrow P_\mu ^\nu D_\nu$$
Since we do not have any replacement rules for symmetric space derivatives of $a_i$ or undifferentiated $a_i$, it is important that in this leading term in $\omega$ expansion, $a_i$ will enter only through $f_{ij}$ and/ or its derivatives

we could see this in the following way.
 The scalars must be invariant under any coordinate transformation and therefore in particular the transformation of the form 
$$t\rightarrow t' = t + f(x_i)$$
Under this coordinate transformation the change in the background metric will also admit an $\omega$ expansion and we know that $a_i$ is the only metric function that will transform even at $\omega\rightarrow0$ limit.

We also know that finally in $S$, the net effect of all such changes should mutually cancel order by order in $\omega$ expansion. However, there would be no candidate to cancel the change in the leading term in $S$, generated due to the transformation of $a_i$ at  order ${\cal O}(\omega^0)$

From here it follows that in the leading term in $\omega$ expansion, $a_i$ could enter only as $f_{ij} = \partial_i a_j -\partial_j a_i$ or  its derivatives or as $\partial_0^k a_i$.

Now these sets of replacement as given in equation \eqref{replacement1}  will produce another covariant scalar $S_{k_i}'$. It is clear that if we again evaluate the covariant difference $(S-S_{k_i}')$ on $\{\hat u^\mu,\hat T,\hat\nu\}$ it will be a scalar of higher order in terms of $\omega$ expansion (though in terms of derivative expansion the order will remain same).

Next we shall apply the same process of replacement for the leading term (in $\omega$ expansion) in this difference $(S-S_{k_i}')$. We shall keep repeating this procedure. Since $S$ is a term of fixed order in derivative expansion, the total number of derivatives will remain constant at each step (only space derivatives will change to time derivatives) and therefore this process must stop at some point.

At the end of this process we shall get a new scalar $S'$ which has the desired form and also reduces to $S$ on $\{\hat u^\mu, \hat T, \hat\nu\}$ in the background with slow time dependence.
This implies 
$$ S = S' + \text{terms that identically vanish on $\{\hat u^\mu, \hat T, \hat\nu\}$ }$$
By construction both $S$ ans $S'$ are of order $n$ in derivative expansion.

Suppose $\{S_i\}$ is the basis of independent scalars at order $n$.
On $\{\hat u^\mu, \hat T, \hat\nu\}$ each $S_i$ must evaluates to some different functions of the background\footnote{This we can see by induction. It is true at first order in derivative expansion. Next we assume that all orders upto $(n-1)$ every independent term evaluates to some different functions of the background. Independent terms at order $n$ could be written either as product of lower order terms or in a form where a symmetric string of $(n-1)$ $D_\mu$ operators (appropriately contracted) acts on a first order term. In either case they will evaluate to different functions of background because of our starting assumption.} and therefore no linear combination of $\{S_i\}$s could vanish on $\{\hat u^\mu, \hat T, \hat\nu\}$. 
Hence it follows that $$S = S'$$
In other words we can always rewrite any dissipative scalar in a form where, in atleast one factor, the $D_\mu $ operator repeatedly acts (with appropriate contractions) on the elements of ${\mathfrak D}$. 

Now we can rearrange such terms further to recast them in the form as given in equation \eqref{vvint}. We need to move the outermost derivatives one by one  by adding total derivative pieces to it. See equation \eqref{exstep} or \eqref{ex:statep} for an example.


\section{Derivation for some  equations}\label{derivation}
In this section we shall derive the important equations that we have used.
\subsection{Divergence of $\hat J^\mu_{can}$}
Here we shall derive equations \eqref{candiv}, \eqref{lasttwo} and \eqref{finalcandiv}.

First we shall manipulate the stress tensor conservation equation in presence of external electromagnetic field.
\begin{equation}\label{eq:deriv1}
\begin{split}
&\frac{\hat u_\nu}{\hat T}\bar\nabla_\mu T^{\mu\nu}= \frac{\hat u_\mu}{\hat T} F^{\mu\nu} J_\nu\\
\Rightarrow &-\frac{1}{\hat T}\left[(\hat E + \hat P) \hat \Theta + \hat u^\mu \partial_\mu \hat E\right] + \bar\nabla_\mu \left(\frac{\tilde\pi^{\mu\nu}\hat u_\nu}{\hat T}\right) = \tilde \pi^{\mu\nu}\bar\nabla_\mu \left(\frac{\hat u_\nu}{\hat T}\right)  -\frac{E_\mu}{\hat T}\tilde j^\mu
\end{split}
\end{equation}
Similarly we can manipulate the equation of current conservation.
\begin{equation}\label{eq:deriv2}
\begin{split}
0=&~\hat \nu\bar\nabla_\mu J^\mu = \hat\nu\bar \nabla_\mu (\hat Q \hat u^\mu + j^\mu)\\
\Rightarrow&~\hat\nu\left[\hat Q \hat \Theta + \hat u^\mu\partial_\mu\hat Q\right] +\bar\nabla_\mu (\hat\nu \tilde j^\mu) = \tilde j^\mu\partial_\mu\hat\nu
\end{split}
\end{equation}
Both in equations \eqref{eq:deriv1} and \eqref{eq:deriv2} we have used equation \eqref{stresscurrent2} for the decomposition of the stress tensor and current.
\newline
Adding equation \eqref{eq:deriv1} and \eqref{eq:deriv2} we get
\begin{equation}\label{eq:deriv3}
\begin{split}
&-\frac{\hat\Theta}{\hat T}\left[(\hat E +\hat P) -\hat \mu \hat Q\right] -\frac{\hat u^\mu}{\hat T}\left[\partial_\mu\hat E -\hat\mu \partial_\mu \hat Q\right] + \bar\nabla_\mu \left(\frac{\tilde\pi^{\mu\nu}\hat u_\nu}{\hat T}\right)+\bar\nabla_\mu (\hat\nu \tilde j^\mu)\\
=& -\tilde j^\mu \left(\frac{E_\mu}{\hat T} - \partial_\mu \hat \nu\right)+\tilde \pi^{\mu\nu}\bar\nabla_\mu \left(\frac{\hat u_\nu}{\hat T}\right) 
\end{split}
\end{equation}
Now we shall use thermodynamics.
\begin{equation}\label{eq:deriv4}
d\hat E = \hat T d s+ \hat\mu d\hat Q,~~\hat E + \hat P = \hat T s + \hat\mu \hat Q
\end{equation}
We could rewrite the first two terms in equation \eqref{eq:deriv3} as 
\begin{equation}\label{eq:deriv5}
-\frac{\hat\Theta}{\hat T}\left[(\hat E +\hat P) -\hat \mu \hat Q\right] -\frac{\hat u^\mu}{\hat T}\left[\partial_\mu\hat E -\hat\mu \partial_\mu \hat Q\right]  = -\bar\nabla_\mu (s \hat u^\mu)
\end{equation}
Substituting equation \eqref{eq:deriv5} in equation \eqref{eq:deriv3} we get the following.
\begin{equation}\label{eq:new0}
\bar\nabla_\mu\left[s \hat u^\mu -\left(\frac{\tilde\pi^{\mu\nu}\hat u_\nu}{\hat T}\right)-(\hat\nu \tilde j^\mu)\right] =\tilde j^\mu \left(\frac{E_\mu}{\hat T} - \partial_\mu \hat \nu\right)-\tilde \pi^{\mu\nu}\bar\nabla_\mu \left(\frac{\hat u_\nu}{\hat T}\right) 
\end{equation}
Now we shall rewrite $\bar\nabla_\mu\left[s \hat u^\mu -\left(\frac{\tilde\pi^{\mu\nu}\hat u_\nu}{\hat T}\right)-(\hat\nu \tilde j^\mu)\right]$ as 

\begin{equation}\label{eq:deriv7}
\begin{split}
&\bar\nabla_\mu\left[s \hat u^\mu -\left(\frac{\tilde\pi^{\mu\nu}\hat u_\nu}{\hat T}\right)-(\hat\nu \tilde j^\mu)\right]\\
= &\frac{1}{\sqrt{G}}\partial_0\left[\sqrt{G}\left(s \hat u^0 -\frac{\tilde\pi^0_\nu\hat u^\nu}{\hat T}-\hat\nu \tilde j^0\right)\right] - \frac{1}{\sqrt{G}}\partial_i\left[\sqrt{G}\left(\frac{\tilde\pi^i_\nu\hat u^\nu}{\hat T}+\hat\nu \tilde j^i\right)\right] 
\end{split}
\end{equation}
Substituting equation \eqref{eq:deriv7} in equation \eqref{eq:new0}, then multiplying both sides of equation \eqref{eq:new0} by $\sqrt{G}$ and integrating over space, we arrive at equation \eqref{candiv}.

To derive equation \eqref{lasttwo} we shall use the following decomposition for the symmetrized derivative of $\left(\frac{\hat u^\nu}{\hat T}\right)$. 
\begin{equation}\label{eq:deriv6}
\begin{split}
\frac{1}{2}\left[\bar\nabla_\mu\left(\frac{\hat u_\nu}{\hat T}\right) +\bar\nabla_\nu\left(\frac{\hat u_\mu}{\hat T}\right)\right] = \frac{\hat\sigma_{\mu\nu}}{\hat T} + \frac{\hat \Theta}{3\hat T}\hat P_{\mu\nu} + \frac{\hat u_\mu\hat u_\nu (\hat u^a\partial_a \hat T)}{\hat T^2}-\frac{1}{2\hat T}\left(\hat u_\mu \hat h_\nu +\hat u_\nu\hat h_\mu\right)
\end{split}
\end{equation}
Where $\hat\sigma_{\mu\nu},~\hat\Theta,~\hat h_\mu$ and $\hat v_\mu$ are defined in equations \eqref{notprofhat}.

Combining equation \eqref{eq:deriv7} and equation \eqref{eq:deriv6} we get the following.
\begin{equation}\label{eq:new1}
\begin{split}
&\frac{1}{\sqrt{G}}\partial_0\left[\sqrt{G}\left(s \hat u^0 -\frac{\tilde\pi^0_\nu\hat u^\nu}{\hat T}-\hat\nu \tilde j^0\right)\right] - \frac{1}{\sqrt{G}}\partial_i\left[\sqrt{G}\left(\frac{\tilde\pi^i_\nu\hat u^\nu}{\hat T}+\hat\nu \tilde j^i\right)\right]\\
&=( \tilde j^\mu \hat u_\mu)(\hat u.\partial\hat \nu) -\left(\frac{\hat u_\mu \hat u_\nu  \tilde \pi^{\mu\nu}}{T^2} \right)(\hat u.\partial \hat T) -\left(\frac{\hat P_{\mu\nu}\tilde \pi^{\mu\nu}}{3\hat T} \right)\hat \Theta\\
 &~~+\hat v_\mu \tilde j^\mu + \left(\frac{\hat u_\nu \tilde \pi^{\mu\nu}}{\hat T}\right)\hat h_\mu -\left(\frac{\tilde \pi^{\mu\nu}}{\hat T}\right)\hat \sigma_{\mu\nu}
\end{split}
\end{equation}

Now by construction 
$$\hat u^\mu\hat h_\mu = \hat u^\mu\hat v_\mu =\hat u^\mu\hat \sigma_{\mu\nu} =0~\Rightarrow \hat h_0 =\hat v_0 = \hat\sigma_{0\mu} =0$$
Therefore 
\begin{equation}\label{simpp}
\begin{split}
&\tilde\pi^{\mu\nu}\hat\sigma_{\mu\nu} = \hat P_\mu^\alpha \hat P_\nu^\beta\tilde\pi^{\mu\nu}\hat\sigma_{\alpha\beta}=\hat P_\mu^i \hat P_\nu^j\tilde\pi^{\mu\nu}\hat \sigma_{ij},~~\tilde\pi^{\mu\nu}P_{\mu\nu} =\tilde\pi^{ij} \hat P_{ij}\\
&\hat u^\nu\tilde\pi^\mu_\nu \hat h_\mu =\hat u^\nu\tilde\pi^\mu_\nu \hat P_\mu ^\alpha \hat h_\alpha = \hat u^\nu\tilde\pi^\mu_\nu \hat P_\mu ^i\hat h_i\\
&\tilde j^\mu\hat v_\mu =\tilde j^\mu \hat P_\mu ^\alpha \hat v_\alpha =\tilde j^\mu \hat P_\mu ^i\hat v_i\\
\end{split}
\end{equation}
 
 Substituting equation \eqref{simpp} in the RHS of equation \eqref{eq:new1} we arrive at equation \eqref{lasttwo} and finally equation \eqref{finalcandiv}.

Equation \eqref{ent:divcan} can be derived in the same way as we have derived equation \eqref{candiv}, we have to just remove all the `$hat$'s from the fluid variables and `$tilde$'s from the stress tensor and current. 
\subsection{Explicit expressions for the hatted quantities}
Here we shall derive the equations \eqref{explicit}. In equations \eqref{explicit} we have the explicit expressions for $\hat \Theta,~\hat\sigma_{\mu\nu},~\hat h_\mu,~\hat v_\mu,~(\hat u.\partial)\hat T$ and $(\hat u.\partial)\hat\nu$. The last two are very simple and we do not need any derivation for them. for the first four we have to do a bit of computation. The first point to note is that all them vanish in equilibrium and therefore their values must be proportional to $\partial_0\Phi$ where $\Phi$ collectively denotes all the metric functions and the gauge fields.

First we shall list all the values for all the Christoffel symbol for the background metric $G_{\mu\nu}$ as given in equation \eqref{equinot}.
\begin{equation}\label{christ}
\begin{split}
\tilde \Gamma^0_{00} &=-e^{2\sigma}(a.\partial)\sigma +a^i\partial_0\left(e^{2\sigma}a_i\right)-\frac{1}{2}\left(a^2 -e^{2\sigma}\right)\partial_0(e^{2\sigma})\\
\tilde\Gamma^i_{00} &=e^{2\sigma}\partial^i\sigma +\frac{a^i}{2}\partial_0\left(e^{2\sigma}\right)-g^{ij}\partial_0\left(e^{2\sigma}a_j\right)\\
\tilde\Gamma^0_{i0} &= \partial_i\sigma -e^{2\sigma}a_i(a.\partial)\sigma  + \frac{e^{2\sigma}a^kf_{ik}}{2} - \left(\frac{a^j}{2}\right)\partial_0\left(g_{ij} - e^{2\sigma}a_ia_j\right)\\
\tilde \Gamma^i_{j0} &=e^{2\sigma}g^{ik}\left(-\frac{1}{2}f_{jk} + a_j\partial_k\sigma\right) +\frac{g^{ik}}{2}\partial_0\left(g_{kj} - e^{2\sigma} a_k a_j\right) \\
\tilde\Gamma^0_{ij} &=-a_k\Gamma^k_{ij} +\frac{e^{2\sigma}}{2}\bigg[a_j a^k\partial_i a_k+a_ia^k\partial_ja_k\bigg]
-\frac{1}{2}(a.\partial)(e^{2\sigma}a_i a_j)\\
&+\frac{e^{-2\sigma}}{2}\bigg[ \partial_i(e^{2\sigma} a_j) +\partial_j(e^{2\sigma} a_i)\bigg]-\frac{1}{2}\left(a^2 -e^{-2\sigma}\right)\partial_0\left(g_{ij} - e^{2\sigma}a_ia_j\right)\\
\tilde\Gamma^k_{ij} &= \Gamma^k_{ij} -\frac{e^{2\sigma}}{2}g^{km}\bigg[a_j\partial_i a_m + a_i\partial_j a_m\bigg] +\frac{1}{2}\partial^k(e^{2\sigma}a_ia_j)
+\frac{a^k}{2}\partial_0\left(g_{ij} - e^{2\sigma}a_ia_j\right)\\
\end{split}
\end{equation}
Here $\partial^j$ denotes $g^{jk}\partial_k$ and $\Gamma^k_{ij}$ 's are the Christoffel symbols for the metric $g_{ij}$.

The expression for electric field is as follows.
\begin{equation}\label{elect}
\begin{split}
E^i = {{\cal F}^i}_\mu \hat u^\mu &=e^{-\sigma}g^{ij}\bigg[\partial_j A_0 - \partial_0(A_j + a_j A_0)\bigg]\\
\end{split}
\end{equation}
Now by construction 
$$\hat u^\mu\hat h_\mu = \hat u^\mu\hat v_\mu =\hat u^\mu\hat \sigma_{\mu\nu} =0~\Rightarrow \hat h_0 =\hat v_0 = \hat\sigma_{0\mu} =0$$
This implies that following.
\begin{equation}\label{anekpare}
\begin{split}
&\hat h^i= G^{i0}\hat h_0 + G^{ij}\hat h_j = g^{ij} \hat h_j\Rightarrow \hat h_i = g_{ij} \hat h^j\\
&\hat v^i= G^{i0}\hat v_0 + G^{ij}\hat v_j = g^{ij} \hat v_j\Rightarrow \hat v_i = g_{ij} \hat v^j\\
&\hat\sigma^{ij} = G^{i\mu}G^{j\nu}\hat\sigma_{\mu\nu}=g^{il}g^{jm}\hat\sigma_{lm} \Rightarrow\hat\sigma_{ij}= g_{il}~ g_{jm} \sigma^{lm}
\end{split}
\end{equation}
Also from the fact that $\hat h_\mu$, $\hat v_\mu$ and $\hat\sigma_{\mu\nu}$ are perpendicular to $\hat u_\mu$ it follows that
\begin{equation}\label{anekpare2}
\begin{split}
&\hat h^0 = -a_i\hat h^i,~~\hat v^0 = -a_i\hat v^i\\
&\hat\sigma^{00} = a_i a_j\hat\sigma^{ij},~~\hat\sigma^{0k} = a_i \hat\sigma^{ik}
\end{split}
\end{equation}
So it is only the space components of $\hat h^\mu$, $\hat v^\mu$ and $\hat\sigma^{\mu\nu}$ that we have to compute explicitly. Time components are related to the space-components through equation \eqref{anekpare2}. 

Now we shall use the definitions of $v^\mu$, $h^\mu$ and $\sigma^{\mu\nu}$ as given in equation \eqref{notprofhat} and shall plug in the expressions of Christoffel symbols and electric field as derived in equation \eqref{christ} and \eqref{elect}. This will give us the equation \eqref{explicit}. For this derivation it is enough to track only those terms that has one explicit factor of $\partial_0$. We already know that the terms without any $\partial_0$ factor must cancel among themselves as all these terms vanish in equilibrium. and equation

\subsection{Divergence of the explicit entropy current constructed in the example}
In this subsection we shall derive equation \eqref{ex:prev}.
The explicit expression for the entropy current is as follows.
\begin{equation}\label{app:ex1}
\begin{split}
J^\mu = \left[s  u^\mu-   \frac{ u^\nu\pi^\mu_\nu}{ T}\right]+K\left[ {\cal G}^{\alpha\beta}(D_\alpha  T)(D_\beta  T)  u^\mu-2{\cal G}^{\mu\beta} (\partial_{\beta}  T)( u^\mu\partial_\mu T)\right] +\tilde S^\mu
\end{split}
\end{equation}
The first term in the square bracket is the canonical part of the entropy current and its divergence we have computed in the previous sections. The final expression for the divergence of the most general canonical entropy current is given in equation \eqref{ent:divcan} where we have to substitute the expression for the stress tensor as given in equation \eqref{ex:expl} (In this example the current vanishes). 
\newline
Divergence of the last term has also been computed for a special case in section (\ref{sec:example}). Here we shall derive the divergence of the second term in equation \eqref{app:ex1}.
\begin{equation}\label{app:ex2}
\begin{split}
&D_\mu \left[ K{\cal G}^{\alpha\beta}(D_\alpha  T)(D_\beta  T)  u^\mu-2K{\cal G}^{\mu\beta} (\partial_{\beta}  T)( u^\mu\partial_\mu T)\right]\\
=&~K\left[(D_\mu T)(D^\mu T)\Theta + 2(D^\mu T)(u.D)(D_\mu T) 
- 2(D^2 T) (u.D T) - 2 (D^\mu T) D_\mu (u.D T)\right]\\
&+ (DT)^2 (u.D K) -2 (D_\alpha K) (D^\alpha T) (u.DT)\\
=&~K\left[(D_\mu T)(D^\mu T)\Theta - 2 (D^\mu T) (D_\mu u^\alpha)(D_\alpha T) - 2 (D^2 T) (u.D T)\right]\\
&+ (DT)^2 (u.D K) -2 (D_\alpha K) (D^\alpha T) (u.DT)\\
=&~K\left[(D_\mu T)(D^\mu T)\Theta - 2 (D^\mu T) \left(\sigma_{\mu\alpha} + \frac{\Theta}{3} P_{\mu\alpha}- u_\mu {\mathfrak a}_\alpha\right)(D^\alpha T) - 2 (D^2 T) (u.D T)\right]\\
&+ (DT)^2 (u.D K) -2 (D_\alpha K) (D^\alpha T) (u.DT)\\
=&~K\left[\frac{\Theta}{3} P^{\mu\nu} (D_\mu T) (D_\nu T) - (u. DT)^2\Theta - 2(D_\mu T)(D_\nu T) \sigma^{\mu\nu} - 2 (D^2 T - {\mathfrak a}.DT)(u.DT)\right]\\
&+ (DT)^2 (u.D K) -2 (D_\alpha K) (D^\alpha T) (u.DT)\\
=&~K\left[\frac{\Theta}{3} P^{\mu\nu} (D_\mu T) (D_\nu T) - (u. DT)^2\Theta - 2(D_\mu T)(D_\nu T) \sigma^{\mu\nu} - 2 (D^2 T - {\mathfrak a}.DT)(u.DT)\right]\\
&+ \frac{dK}{dT}(DT)^2 (u.D T) -2\frac{dK}{dT} (D_\alpha T) (D^\alpha T) (u.DT)\\
=&~K\left[\frac{\Theta}{3} P^{\mu\nu} (D_\mu T) (D_\nu T) - (u. DT)^2\Theta - 2(D_\mu T)(D_\nu T) \sigma^{\mu\nu} - 2 (D^2 T - {\mathfrak a}.DT)(u.DT)\right]\\
&-\frac{dK}{dT}(DT)^2 (u.D T)
\end{split}
\end{equation}
Here by $(DT)^2$ we mean $(D_\alpha T )(D^\alpha T)$ and $(u.D)$ denotes the operator $(u^\alpha D_\alpha)$.
\newline
In the third line we have used the following decomposition.
$$D_\mu u_\nu = \sigma_{\mu\nu} + \frac{\Theta}{3}P_{\mu\nu} + \omega_{\mu\nu} - u_\mu {\mathfrak a}_\nu$$
where
\begin{equation*}
\begin{split}
&P^{\mu\nu} = {\cal G}^{\mu\nu} + u^\mu u^\nu\\
 &\Theta = D_\mu u^\mu,~~~
{ \mathfrak a}_\mu = (u.D)u_\mu\\
 &\sigma_{\mu\nu} = P_\mu^{\alpha} P_\nu^\beta\left[\frac{D_\alpha u_\beta +D_\beta u_\alpha }{2}- \frac{\Theta}{3} {\cal G}_{\alpha\beta}\right]\\
 &\omega_{\mu\nu} = P_\mu^{\alpha} P_\nu^\beta\left[\frac{D_\alpha u_\beta -D_\beta u_\alpha }{2}\right]
\end{split}
\end{equation*}
Combining equation \eqref{app:ex2} with the divergence of the canonical entropy part (equation\eqref{ent:divcan} with $j^\mu\rightarrow0$) and using equation \eqref{ex:expl} for the explicit form of $\pi^{\mu\nu}$, we arrive at equation \eqref{ex:prev}.

\section{Existence of a near equilibrium solution}\label{app:solution}
Throughout this note we have assumed that there exists at least one solution to the fluid equations so that the stress tensor and the current evaluated on that solution admit a power series expansion in $\omega$ around $(\omega =0)$ . We have not been able to to prove the existence of such solution for a general set of hydrodynamic equations. Instead, here we shall study some very simple equations. These are the equations that motivated this assumption we are using throughout.
\subsection{SHM with equilibrium slowly shifting with time}
Suppose we are to study the following equation.
\begin{equation}\label{SHM1}
\begin{split}
&\frac{d^2x}{dt^2} + \alpha\frac{dx}{dt} + k [x -a(t)] =0\\
\text{where}~~&a(t) = \bar a ~e^{i\omega t}
\end{split}
\end{equation}
If $\omega=0$ then the only time-independent solution would have been $x = \bar a$. Here $\bar a$ is equilibrium position. We would like to see whether for small $\omega$ we could have solution of the form $$x =[ \bar a  + {\cal O}(\omega)]e^{i\omega t}$$
Now $a(t)$ is a forcing term for the equation and the following is one particular solution for the system.
 \begin{equation}\label{SHM2}
 \begin{split}
 x_p(t) = \left(\frac{k}{-\omega^2 + i\alpha\omega +k}\right)\bar a e^{i\omega t}
 \end{split}
\end{equation}
This particular solution admits an power series expansion in $\omega$.
\begin{equation}\label{SHM3}
 \begin{split}
 x_p(t) = \left[1 -\left(\frac{i\alpha\omega}{k}\right) + \frac{\omega^2}{k^2}(k- \alpha^2) + {\cal O}(\omega^3)\right]\bar a e^{i\omega t}
 \end{split}
\end{equation}
Homogeneous solution has the following form
\begin{equation}\label{SHM4}
 \begin{split}
 x_h(t) = e^{-\alpha t}\left(Ae^{t\sqrt{\alpha^2-4k} } + Be^{-t\sqrt{\alpha^2-4k}  }\right)
 \end{split}
\end{equation}
The full solution is $~~~~x(t) = x_p(t) + x_h(t)$
\newline
 If $\alpha<0$, the homogeneous part will diverge at large time and will take the solution far from its equilibrium value. On other hand if $(\alpha>0) $ and $(\alpha^2<4k)$, then the homogeneous part of the solution will eventually decay and $x_p(t)$ will be the solution at large time.
\newline
However, irrespective of the value of $\alpha$ we can always choose a very fined tuned initial condition such that both $A$ and $B$ are zero. In such case $x_p(t)$ itself is the solution for all time and it does have the desired form in terms of $\omega$ expansion. 

In this equation the coefficient $\alpha$ is the damping term and therefore analogous to the dissipative coefficients like viscosity ($\eta$) and bulk viscosity ($\zeta$) etc. $a(t)$ is equivalent to the slow time dependence of the equilibrium. The existence of this fine tuned solution for the equation \eqref{SHM1} motivates our assumption about the existence of a fluid solution that admits a power series expansion in terms of $\omega$ around $\omega =0$.

\subsection{Chemical potential in a slowly varying electric field}
Here we shall study another simple equation which has a little more resemblance with fluid equations than the previous one.

Suppose we are studying the current conservation equation where chemical potential is the only fluid variable and $A_0$ is the only background component that has non trivial dependence on space-time.

Suppose the constitutive relation for the current is the following.
\begin{equation}\label{CHM1}
\begin{split}
&C^0 =q(\nu),~~C^i=\Delta \left(\frac{E^i}{T} - \partial^i\nu\right)\\
\text{where}&~~\Delta = \text{constant},~~T = \text{constant},~~\frac{E^i}{T} = \frac{\partial^iA_0}{T}
\end{split}
\end{equation}
Current conservation equation reads as follows.
\begin{equation}\label{CHM2}
\begin{split}
&\left(\frac{dq}{d\nu}\right)\partial_0\nu + \Delta \left(\frac{\partial^2A_0}{T} - \partial^2\nu\right) =0
\end{split}
\end{equation}
Suppose $A_0(x_i,\omega) = \bar A_0 e^{ik_i x_i + i \omega t}$ and $\omega\ll k_i$ for every $k_i$
\newline
At $\omega =0$ only time-independent solution for the system would be
$$\nu(x_i) = \frac{\bar A_0 e^{ik_i x_i}}{T} = \frac{A_0(x_i)}{T}$$

For non-zero $\omega$ one particular solution to the equation \eqref{CHM2} could be written as 
\begin{equation}\label{CHM3}
\begin{split}
&\nu_p(x_i,t) =\frac{1}{T} \bigg[\frac{\Delta k^2}{\Delta k^2  + i \omega \left(\frac{dq}{d\nu}\right)}\bigg]\bar A_0 e^{ik_i x_i + i \omega t}
\end{split}
\end{equation}
As before this particular solution admits a power series expansion in terms of $\omega$.
\begin{equation}\label{CHM4}
\begin{split}
&\nu_p(x_i,t) =\left(\frac{A_0(x_i,t)}{T} \right)\bigg[1 - \frac{i\omega}{\Delta k^2} \left(\frac{dq}{d\nu}\right) - \frac{\omega^2}{\Delta^2 k^4}\left(\frac{dq}{d\nu}\right)^2 + {\cal O}(\omega^3)\bigg]
\end{split}
\end{equation}
The homogeneous solution is as follows.
\begin{equation}\label{CHM5}
\begin{split}
&\nu_h(x_i,t) =B\exp\left[-t\Delta k^2\left(\frac{dq}{d\nu} \right)^{-1}  \right]e^{i k_i x_i}
\end{split}
\end{equation}
The full solution is $~~~\nu = \nu_h + \nu_p$
 \newline
As before if $\Delta<0$, $\nu_h$ will grow with time and the full solution cannot be written as slowly shifting equilibrium plus small correction. On the other hand if $\Delta>0$ the effect of the homogeneous part will decay with time and at large time the solution is effectively the particular part.

However, theoretically we can always choose an initial condition so that $B=0$.  For this very fine tuned case $\nu_p(x_i,t)$ is a solution for all time. Thus we see that irrespective of the sign of $\Delta$ there exists a solution of the desired form.  
Another point we note here is that in the solution $\nu_p(x_i,t)$, for different powers of $\omega$,  we have different powers of space momenta in the denominator. This indicates that the solution is  non-local in the space direction.

We believe that these qualitative features will be true even for the solutions to the actual fluid equations which are highly non-linear and coupled. But we do not have any proof for this.

\providecommand{\href}[2]{#2}\begingroup\raggedright\endgroup

\end{document}